\documentclass[aps,prb,twocolumn,groupedaddress]{revtex4}
\usepackage{graphicx}
\usepackage{dcolumn}
\usepackage{bm}
\usepackage{amsmath,amsfonts,amssymb,amsthm,verbatim}
\usepackage[ansinew]{inputenc}
\usepackage[usenames]{pstricks}
\usepackage{color}
\usepackage{epsfig}
\usepackage{MnSymbol}
\usepackage{bbold}

\newcommand{\be}{\begin{equation}}
\newcommand{\ee}{\end{equation}}
\newcommand{\bea}{\begin{eqnarray}}
\newcommand{\eea}{\end{eqnarray}}

\newcommand{\bi}{\begin{itemize}}
\newcommand{\ei}{\end{itemize}}
\newcommand{\bc}{\begin{center}}
\newcommand{\ec}{\end{center}}

\begin{document}

\title{Transport signatures of symmetry protection in 1D Floquet topological insulators}

\author{Oleksandr Balabanov}
\affiliation{Department of Physics, University of Gothenburg, SE 412 96 Gothenburg, Sweden}

\author{Henrik Johannesson}
\affiliation{Department of Physics, University of Gothenburg, SE 412 96 Gothenburg, Sweden}

\begin{abstract}

Time-periodic external drives have emerged as a powerful tool to artificially create topological phases of matter. Prime examples are Floquet topological insulators (FTIs), where a gapped bulk supports in-gap edge states, protected against symmetry-preserving local perturbations. Similar to an ordinary static topological insulator, the robustness of an edge state in a one-dimensional (1D) FTI shows up as a pinning of its quasienergy level, but now inside one of two distinct bulk gaps. Here we propose a scheme for probing this unique feature by observing transport characteristics of a 1D finite-sized FTI  attached to external leads. We present predictions for transmission spectra using a nonequilibrium Green's function approach. Our analysis covers FTIs with time-independent and periodically driven boundary perturbations which either preserve or break the protecting chiral symmetry.  \\

\end{abstract}

\maketitle

\section{Introduction}

Periodically driven quantum systems $-$ {\em quantum Floquet matter} \cite{Bukov,MoessnerSondhi} $-$ 
exhibit a number of intriguing phenomena, many of which are not seen in equilibrium. 
\text{Notable} examples are the Floquet topological insulators (FTIs), obtained by applying an external time-periodic field which couples to
some weakly correlated (or noninteracting) degrees of freedom of an appropriate system, by this producing an insulating phase characterized by a set of nonzero topological invariants \cite{Kitagawa,Lindner,Kitagawa2,Gu,Rudner,Asboth,Carpentier,Nathan,Fruchart,Roy,Schuster}.

Similar to an ordinary topological band insulator \cite{Chiu}, an FTI hosts robust edge states, originating from the 
nontrivial topology of the states in the bulk (``bulk-boundary correspondence" \cite{Rhim}) and protected against local perturbations provided that these perturbations respect some 
relevant underlying symmetry. Different from a static topological insulator, however, the nontrivial topology of an FTI refers to the nonequilibrium states excited by the periodic drive. 
This opens up for the appearance of an additional branch of symmetry-protected edge states \cite{ }: Given a driving frequency $\Omega$, boundary levels may now appear not 
only in the bulk gap around zero quasienergy (playing the role of ``energy" in the driven system), but also in the ``anomalous" quasienergy gap at $\pm \hbar\Omega/2\,$ \cite{Rudner,Carpentier,Roy}.

Band structures and certain other features of FTIs have been captured in experiments with shaken optical lattices \cite{Hauke,Jotzu,Aidelsburger,Fujiwara} and photo-induced states \cite{Rechtsman,Wang,Cardano}. In a recent breakthrough \cite{McIver}, heralding high-precision transport experiments on electronic FTIs, the predicted quantized Hall conductance of monolayer graphene driven by circularly polarized light \cite{Kitagawa2,Torres,Usaj,Dehgani,Mikami} was also measured. The quantization of longitudinal edge state transport in FTIs (or absence of such) has been recently addressed as well\,\cite{Gu,Torres,KunduFertigSeradjeh, Farrell1,Farrell2, Kundu}, however, until now only theoretically - waiting to be confirmed in experiments. 
Other topics on conductance quantization in FTIs, including the use of a ``scattering matrix invariant" \cite{Fulga}, have also been proposed. In parallel developments, theoretical works have uncovered unusual dc conductance scaling and other anomalous properties of edge state transport in irradiated graphene, in ribbon \cite{Gu} and cylindrical geometries \cite{Dehgani,Dehgani2,Dehgani3}. Transport in a semi-infinite 1D FTI was analyzed in Ref.~\onlinecite{Ruocco}, using a model of a periodically driven dimer array. It was predicted that with the array contacted to an external lead, a current will respond continuously to variations of the driving-field amplitude if there is a localized topological edge state adjacent to the contact. Other theoretical efforts include studies of how bulk disorder \cite{Titum,KRBL} and energy and momentum relaxation \cite{Seetharam,Esin} influence 2D FTI edge state transport,
with yet other topics addressed in Refs. \onlinecite{Yap1,Yap2,ZhouGong,HuamanUsaj,Goldman}.  All these works testify that the problem of FTI edge transport is highly nontrivial and that additional ideas for experiments and their interpretations are called for.

In this paper we contribute to this effort by suggesting a blueprint for probing Floquet symmetry protection of the topological edge states in 1D FTIs using transport measurements. In short, we suggest to use a 1D topologically insulating mesoscopic structure connected to external leads biased by an applied dc voltage. In contrast to 2D and 3D topological insulators which host linearly dispersing edge states which, ideally, support ballistic transport \cite{Hasan,Qi}, a 1D topological insulator exhibits edge states stuck at its two edges \cite{AsbothBook}. Still, in a {\em finite-sized} system, the hybridization of the edge states may open a channel for coherent transport of electrons, making them share a feature akin to their higher-dimensional relatives. Here we confirm that this property also holds for FTIs, and propose a scheme where measuring a dc current across a 1D finite FTI is expected to yield distinct fingerprints of the symmetry protection of its edge states, both at $\varepsilon=0$ {\em and} $\varepsilon=\hbar \Omega/2$ quasienergies: By employing a nonequilibrium Green's function approach we monitor how the in-gap transmission peaks, present due to hybridization of the symmetry-protected edge states, rapidly disappear under local time-independent {\em and} time-periodic symmetry-breaking perturbations while remaining if the symmetries are maintained. Our results show that there is no significant difference in the degree of symmetry-protection in the normal gap (at $\varepsilon=0$) as compared to the ``anomalous" gap (at $\varepsilon=\hbar \Omega/2$), however, the transmission via edge states in the anomalous gap appear to be more fragile against symmetry-breaking time-periodic perturbations. This happens due to the intrinsic structural difference of the topological edge states at $\varepsilon=0$ and $\varepsilon=\hbar \Omega/2$ described in detail in the main text and Appendices. As a model realization of 1D FTI for producing numerics we consider an array of dimers described by a periodically driven spinless Su-Schrieffer-Heeger (SSH) model \cite{SSH}, which, while being simple, embodies all relevant topological properties.

The paper is organized as follows: In Sec. II we set the stage by introducing a dimer array connected to external leads, and predict the transmission spectra for both undriven (time-independent) and periodically driven (Floquet) systems using Landauer-B\"{u}ttiker theory \cite{Datta} within a nonequilibrium Green's function formalism. While the undriven case is admittedly somewhat trivial, using it as a backdrop is instructive for intuition and interpretation of the Floquet case. This is particularly so since we will take advantage of a 
calculational method recently introduced by one of us \cite{OB2018}, exploiting analogies between well-known time-independent quantities in Landauer-B\"utttiker theory and the corresponding objects defined within the frequency domain of Floquet theory. In Sec.~III we then turn to the central theme of the paper: the symmetry protection of the edge states. Our analysis, again employing a nonequilibrium Green's function approach, covers both time-independent and periodically driven systems 
with time-independent and periodically driven boundary perturbations which either preserve or break the protecting underlying symmetry, here being a chiral symmetry. Sec. IV briefly summarizes our results with a short outlook for future work. Technical details are spelled out in four appendices. 

\section{Transport Through a Finite Topologically Nontrivial System}

As prototype of a 1D topological insulator we take an array of dimers described by a spinless SSH model with Hamiltonian \cite{SSH}
\begin{align} 
\begin{split} 
H  = & \sum_{j} \left( \gamma_1(t) |A, j \rangle \langle B, j| + \gamma_2(t) |B, j-1 \rangle \langle A, j| + \text{H.c.} \right), 
\end{split}
\label{eq:SSH_original}
\end{align}
up to a uniform term determined by an applied gate voltage, see Fig. 1. Here $A, B$ label the two monomers of a dimer, indexed by $j$, and $\gamma_1(t)$ and $\gamma_2(t)$ are time-periodic intracell and intercell hopping amplitudes respectively. We have suppressed the spin degree of freedom in Eq. (\ref{eq:SSH_original}) as it merely produces a factor of two in all results that follow. The spin can always be restored if needed. 

The SSH model serves as a minimal model for a topological insulator.  When there is no external drive, i.e. $\gamma_1(t) = \gamma_{1,0}$ and $\gamma_2(t) = \gamma_{2,0}$ with $\gamma_{1,0}$ and $\gamma_{2,0}$ constants, the model supports two topologically distinct phases, trivial ($|\gamma_1|>|\gamma_2|$) and nontrivial ($|\gamma_1|<|\gamma_2|$), with topological invariants $\nu=0$ and $\nu=1$ respectively \cite{SSHtopo3}. {\em With} a time-periodic external drive, there are in total {\em four} topological phases labeled by {\em two} topological indices, $\nu_0$ and $\nu_{\pi}$. 
As shown in Refs. \onlinecite{Lago}, \onlinecite{OBHJ2017}, which one of the phases is realized depends on the nature of the drive, as well as the relation between its frequency and the magnitude of the hopping amplitudes. By the bulk-boundary correspondence, $\nu_0 \, (\nu_{\pi})$ counts the number of time-periodic edge states in the gap at $0$- $(\hbar \Omega/2)$-quasienergy.

\begin{figure} \centering
    \includegraphics[width=7.5cm,angle=0]{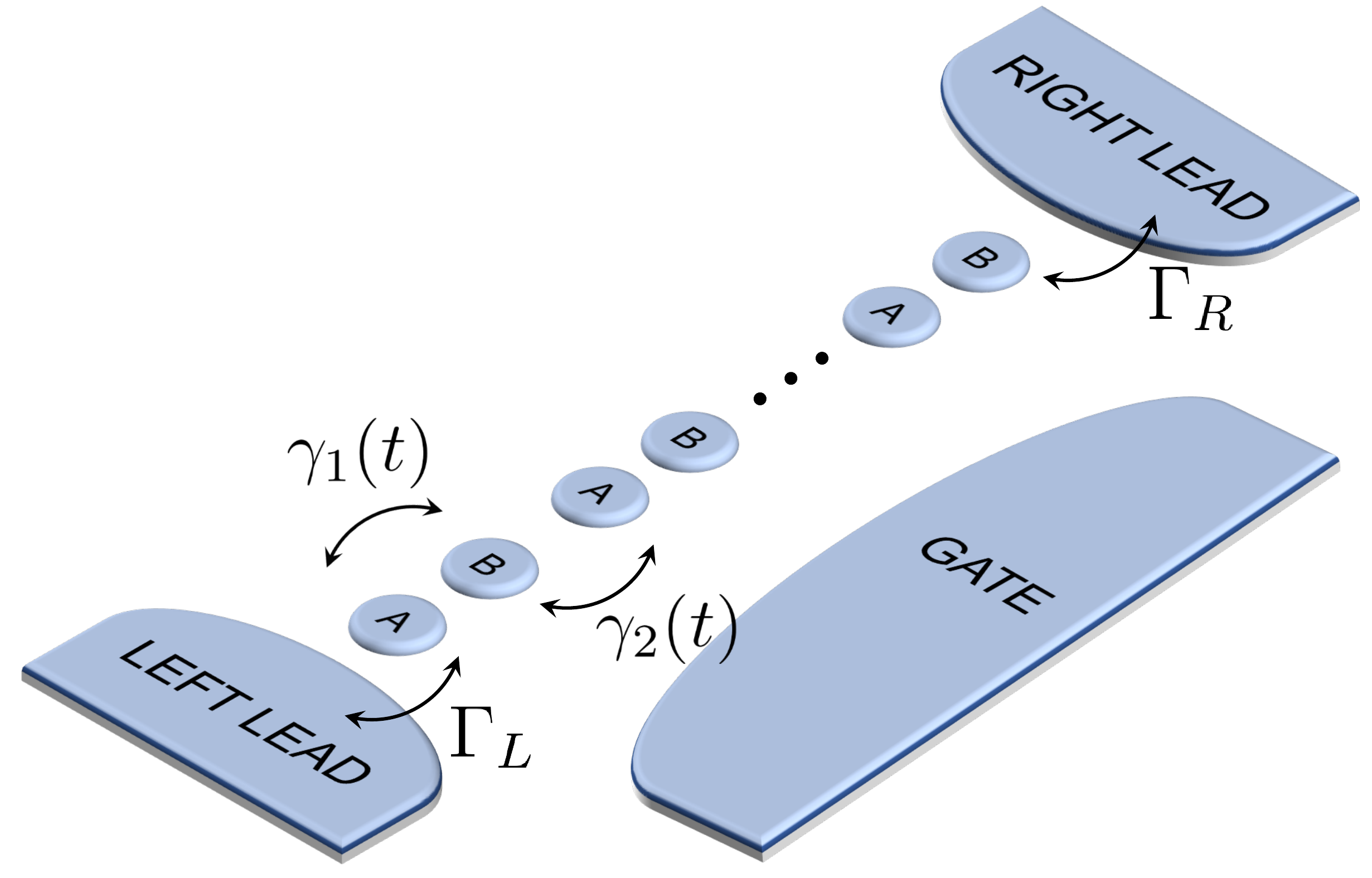}
    \caption{Schematic setup of an array of dimers connected to two leads (with contacts represented by matrices $\Gamma_L$ and $\Gamma_R$). The unit cell consists of two sites, labeled by A and B, and the intracell (intercell) hopping amplitude is $\gamma_1(t)$ ($\gamma_2(t)$).}
     \label{fig1}
\end{figure}

We connect a 1D topological insulator to two external leads by contacts represented by matrices $\Gamma_L$ and $\Gamma_R$ respectively
, see Fig. 1. Applying a gate voltage allows for a shift of the chemical potential of the transport system with respect to the potentials of the leads, where the latter are biased by an applied voltage~$V$. 
 In the thermodynamic limit the topologically nontrivial phase is distinguished by the presence of zero-energy modes localized at the edges. When the length of the topological insulator is finite, these states hybridize and create additional transport channels across the structure. It follows that by attaching leads to the system one expects to observe nonzero transmission peaks even at biases within the bulk gap. Such peaks should gradually disappear with increasing size of the insulator, while being completely absent in a topologically trivial phase. In this way fingerprints of the topological edge states can be established. 

\begin{figure} \centering
    \includegraphics[width=6.5cm,angle=0]{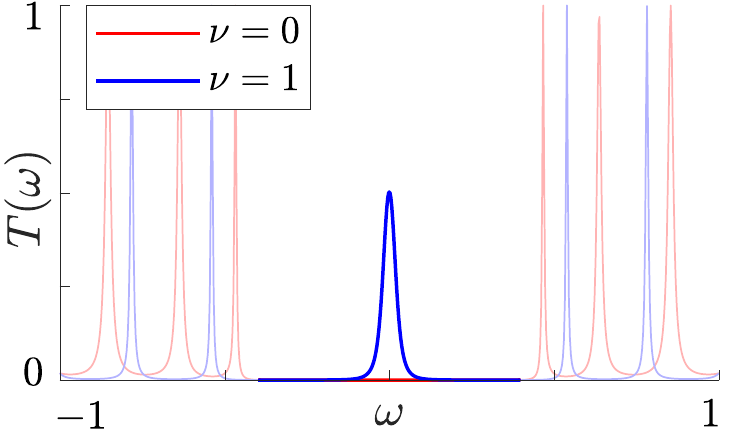}
    \caption{Transmission spectra $T(\omega)$ corresponding to topologically trivial ($\gamma_1 = 1.2$, $\gamma_2 = 0.8$,  red color) and nontrivial ($\gamma_1 = 0.8$, $\gamma_2 = 1.2$, blue color) undriven dimer arrays.
    Here we have considered arrays with 9 dimers, with nonzero contact matrix elements (``broadening functions") $\Gamma_{L,\, A 1} = \Gamma_{R,\, B N} = 0.1$. The transmission spectra corresponding to the in-gap states are illustrated with higher color intensity.}
     \label{fig2}
\end{figure}
\subsection{Transmission Through Undriven Dimer Arrays}
Measurable transport characteristics for an {\em undriven} mesoscopic system, like the dc current and conductance, is easily obtained by employing standard Landauer-B\"{u}ttiker theory within a nonequilibrium Green's function formalism (for details, see Appendix~A). 
The dc current $I_\text{dc}$ across the central region is given by the Landauer-B\"{u}ttiker formula \cite{Datta}
\begin{align}   
\begin{split}
I_\text{dc} =  \frac{e}{h}  \int_{-\infty}^{\infty} \, d\omega\,    \,& T (\omega) \,  [f(\omega-eV_L) - f(\omega-eV_R)]. \\
\end{split}
\label{eq:dc_current}
\end{align}
Here $f(\omega-eV_L)$  and $f(\omega-eV_R)$ 
are Fermi-Dirac distribution functions with $V_L$ and $V_R$ the electrostatic potentials of the left and right leads respectively, with $e$ the electron charge. $T(\omega)$ is the transmission spectrum, being a function of $\Gamma_L, \Gamma_R$, and the electron density of states (cf. Appendix~ A). The spin degree of freedom, if present, is included in $T(\omega)$. We here consider only a small applied bias $V = V_L-V_R$ so that the transmission $T(\omega)$ can be considered to be independent of $V$. In an experiment this corresponds to measurements of the linear conductance, equal to $(e^2/h)T(\omega = 0)$ when at zero temperature. To probe $T(\omega)$ for different energies~$\omega$ one then measures the low-temperature linear conductance at different values of the applied gate voltage, in this way sweeping the experimentally accessible part of the spectrum of the topological insulator.

The transmission spectra corresponding to two topologically distinct finite SSH chains, trivial and nontrivial, are illustrated in Fig.~2. In the topologically nontrivial case (blue color) one identifies a transmission peak at $\omega=0$ as expected from the discussion above, as well as from related work in Ref. \onlinecite{Dong}: In the topologically nontrivial phase the two degenerate 
zero-energy edge states hybridize and open a conduction channel across the array. The width of the peak is proportional to the broadening of the zero-energy boundary levels due to their mutual hybridization as well as to the coupling to the external leads, here taken to be the same for left and right lead, $\Gamma_L = \Gamma_R$.  

It is important to know how the in-gap transmission peak changes with the system size. In this way one may predict when to expect the largest transmission due to the presence of the topological edge states. In Appendix~C we develop a theory from which such scaling can be predicted. The idea is to project the transport problem onto the closest states with zero energy (i.e. the edge states split by their mutual hybridization), and assume that the ``high-energy" states of the bulk do not contribute to the transport at small applied bias. We then calculate the Green's function projected onto the space spanned by these states. The transmission spectrum is then directly retrieved. For the simple case of the dimer array, Eq. (1), we have succeeded to derive a fully analytic expression, however, for a 1D topological insulator with a more complicated Hamiltonian one may first have to perform a numerical diagonalization. For the analytic calculation scheme, see Appendix~C. Here we present only the result: In Fig.~3 we show zero transmission $T(\omega = 0)$ vs. number of dimers used in the topologically nontrivial array. The analytic result (red) agrees well with the numerical data (blue), confirming the prediction derived in Appendix C.

\begin{figure} \centering
    \includegraphics[width=7cm,angle=0]{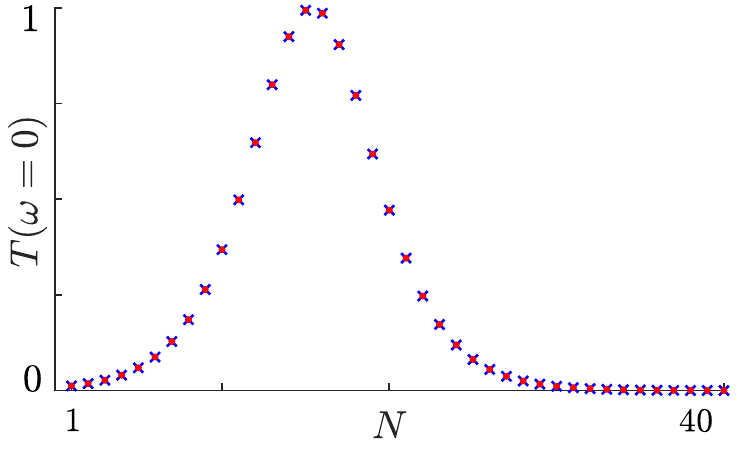}
    \caption{Size-dependence of the transmission at $\omega = 0$ with number of dimers $N$ in an undriven array. Blue crosses represent data from a numerical computation using Green's functions, with the red dots derived from an analytic calculation applying edge state projection. The hopping amplitudes are set to $\gamma_1 = 0.9$ and $\gamma_2 = 1.1$, the broadening functions to $\Gamma_{L,\, A 1} = \Gamma_{R,\, B N} = 0.1$.}
     \label{fig3}
\end{figure}

\subsection{Transmission Through Periodically Driven Dimer Arrays}

An approach similar to the one above can be applied also for distinguishing topologically distinct phases in FTIs. As suggested by this very term, periodically driven quantum systems are conveniently described within the Floquet formalism~\cite{Shirley}. This theory states that there is a complete set of solutions to the time-periodic Schr\"odinger equation of the form $|\psi(t)\rangle = |u(t)\rangle \exp(-i\varepsilon t)$, where $|u(t)\rangle = |u(t+T)\rangle$ and $\varepsilon$ are real constants called quasienergies. Often it is practical to represent time-periodic modes $|u(t)\rangle$ via a Fourier transformation $|u^{(m)}\rangle = 1/T \, \int_0^T \, dt \, \exp(i m \Omega t) |u(t)\rangle$, where $|u^{(m)}\rangle$ is a Floquet component of the state $|u(t)\rangle$ corresponding to the Floquet index $m$. The quasienergies form bands very similar to energy bands of a time-independent spatially periodic system. However, the quasienergies are defined only modulo $\hbar \Omega$ where $\Omega = 2\pi/T$ is the frequency of the drive, and by this induce an additional repetition of the bands. Such quasienergy band structures can be described within topological band theory and topologically distinct phases can be identified similar to the case of time-independent systems~\cite{Kitagawa, Lindner}. However, the fact that the nontrivial topology in an FTI derives from the unitary time evolution leads to the possibility to have symmetry-protected edge states also inside the anomalous gap at $\hbar \Omega/2$. This makes the FTIs different from their static relatives \cite{Rudner,Carpentier,Roy}.

We shall again focus on the dimer array described by Eq.~(1), but now considering the hopping amplitudes $\gamma_1(t) = \gamma_{1,0} - v(t)$ and $\gamma_2(t) = \gamma_{2,0} + v(t)$ with a nonzero time-modulation $v(t) \sim \cos(\Omega t)$. Such arrays represent a {\em harmonically driven SSH model}\,\cite{Lago,OBHJ2017,SeradjehSSH1,SeradjehSSH2,photonic2}, a minimal model of an FTI exhibiting all the relevant topological phases, with possible edge states not only at zero quasienergy but also at $\hbar \Omega/2$.
Both types of edge states create additional transport channels in finite arrays via hybridization and one should therefore be able to see them in conductance measurements. Let us stress that our analysis, to be carried out below, does not rely on the particular model or driving considered; it can be extended to any model of a 1D FTI. Indeed, in an experiment one may look for other, maybe more easily attainable realizations of topologically nontrivial 1D systems.

There is a variety of methods available to calculate periodically driven transport in mesoscopic systems \cite{Platero,Kohler}, including master equations \cite{Bruder}, Floquet scattering matrices
\cite{Moskalets,Kim,ArracheaMoskalets}, and nonequilibrium Green's functions \cite{Jauho,Arrachea,ArracheaMoskalets,Tsuji}. Here we adopt the Green's function approach formulated within Floquet-Sambe space~\cite{OB2018}. 
The advantage of this method compared to other equivalent techniques lies in its simplicity: Expressions for currents and densities essentially replicate well-known time-independent formulas but with time-independent objects replaced by their analogs within Floquet-Sambe theory. In this way the dc (time-averaged) current $I_{\text{dc}}$ takes the form
\begin{align}
\begin{split}
I_{\text{dc}} \! =\! \frac{e}{h} \int_{-\infty}^{\infty} \!d\omega   &   
\,\big(\mathcal{T}_{LR}^{\, (0)} (\omega) f(\omega \!-\!eV_L) 
-  \mathcal{T}_{RL}^{\, (0)} (\omega) f(\omega \!-\!eV_R)\!\big),  \\ 
\end{split}
\label{eq:current_total_FS}
\end{align}
where $\mathcal{T}_{\nu, \, \nu^\prime}^{\, (0)}(\omega)$ denotes the amplitude for a photon-assisted transmission of an electron from lead~$\nu$ to lead~$\nu^\prime$ at energy $\omega$, and $V_\nu$ is the electrostatic potential of lead~$\nu$ ($\nu, \, \nu^\prime = L, R$). Similar to the time-independent case, the spin degree of freedom, if present, is included in $\mathcal{T}_{LR}^{\, (0)}$ and $\mathcal{T}_{RL}^{\, (0)}$. Note that this formula is in agreement with analogous ones derived using other methods, see e.g. Ref. \onlinecite{Arrachea} with a derivation carried out using a Floquet-Keldysh formalism. In contrast to the time-independent case in Eq.~(\ref{eq:dc_current}), Eq.~(\ref{eq:current_total_FS}) is not antisymmetric under exchange of the leads since for a driven system generally $\mathcal{T}_{LR}^{\, (0)} \neq \mathcal{T}_{RL}^{\, (0)}$ \cite{Kohler}. As a consequence, the differential conductance, $G=dI_{\text{dc}}/dV$, depends crucially on the voltage profile across the transport region. In the idealized case where the voltage drop is entirely at the contact between lead $L$ and the central system, $V_L=V$ and $V_R=0$, Eq.~(\ref{eq:current_total_FS}) implies that the zero-temperature linear (small bias $V$) conductance is given by \cite{Fruchart2} 
\begin{equation}
G_{LR} = (e^2/h) \mathcal{T}_{LR}^{\, (0)}(\omega =0),
\end{equation}
with $L \leftrightarrow R$ when the voltage drop is entirely at the contact between lead $R$ and the central system. As before, to access the whole transmission range in $\omega$ one should perform the measurements at various gate biases. In~the more realistic case, with an extended voltage profile across the transport region, the measured zero-temperature linear conductance will receive contributions from both transmission spectra and take a value between $(e^2/h) \mathcal{T}_{LR}^{\, (0)}({\omega})$ and $(e^2/h) \mathcal{T}_{RL}^{\, (0)}({\omega})$. It follows that the topological fingerprints in both transmission spectra, to be uncovered below, should also be present in the realistic case of a linear conductance measurement. A short technical discussion on the computational approach including its numerical implementation is given in Appendix~B. For a more detailed account we refer the reader to Ref.~\onlinecite{OB2018}. 
In what follows we display results only for $\mathcal{T}_{LR}^{\, (0)}(\omega)$, from now on simply denoted by $\mathcal{T}^{\, (0)}(\omega)$, and skip the very analogous discussion of $\mathcal{T}_{RL}^{\, (0)}(\omega)$ to avoid repetition.

\begin{figure} \centering
    \includegraphics[width=8.5cm,angle=0]{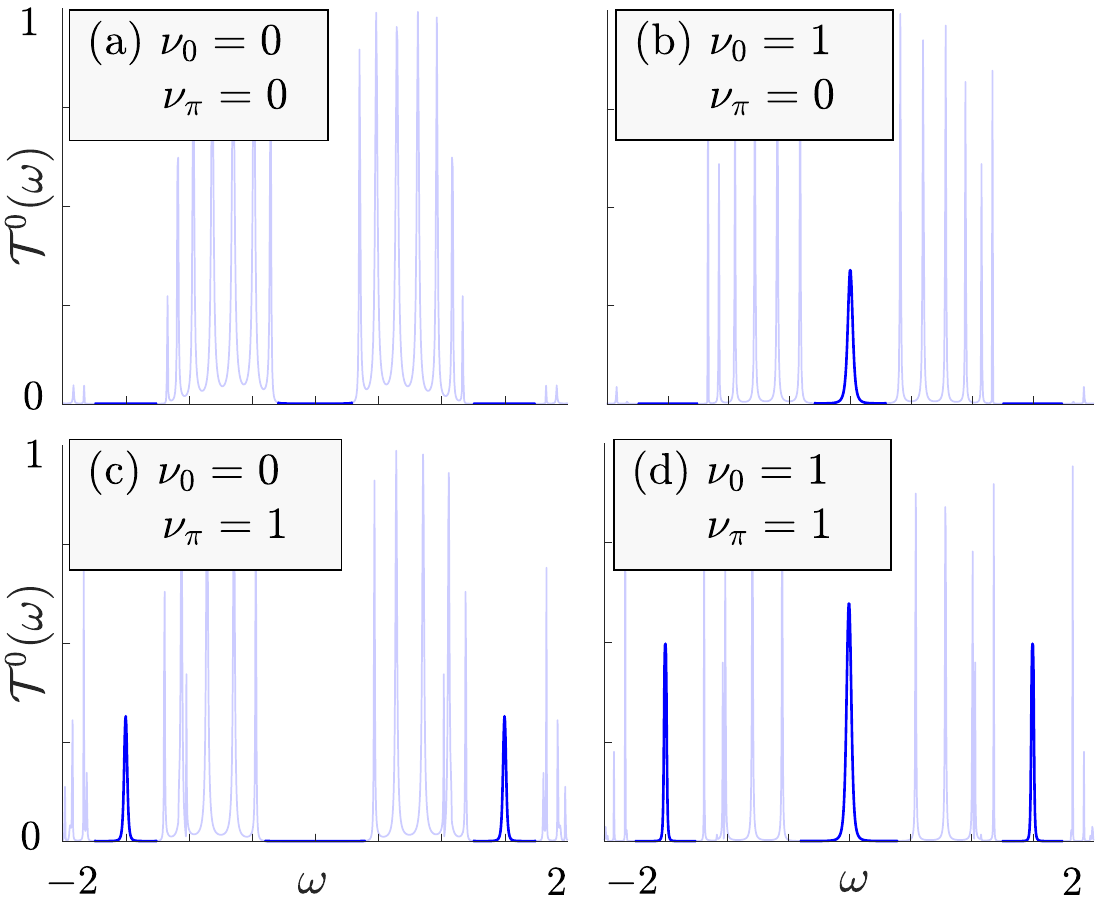}
        \caption{Transmission spectra corresponding to the four different topological phases of the periodically driven arrays with (a) $\gamma_{1,0} = 0.75$, $\gamma_{2,0} = 0.45$; (b) $\gamma_{1,0} = 0.45$, $\gamma_{2,0} = 0.75$; (c) $\gamma_{1,0} = 1.2$, $\gamma_{2,0} = 0.8$; (d) $\gamma_{1,0} = 0.8$, $\gamma_{2,0} = 1.2$. The topological invariants $\nu_0$ and $\nu_\pi$ label the topological class of the corresponding array. In all four cases the arrays consist of 9 dimers, the driving is set to $v(t) = 0.4 \cos (3t)$, and the broadening functions are taken to be $\Gamma_{L,\, A 1}= \Gamma_{R,\, B N} = 0.1$. The transmission spectra corresponding to the in-gap states are illustrated with higher color intensity.}
     \label{fig4}
\end{figure}
\begin{figure} \centering
    \includegraphics[width=8.5cm,angle=0]{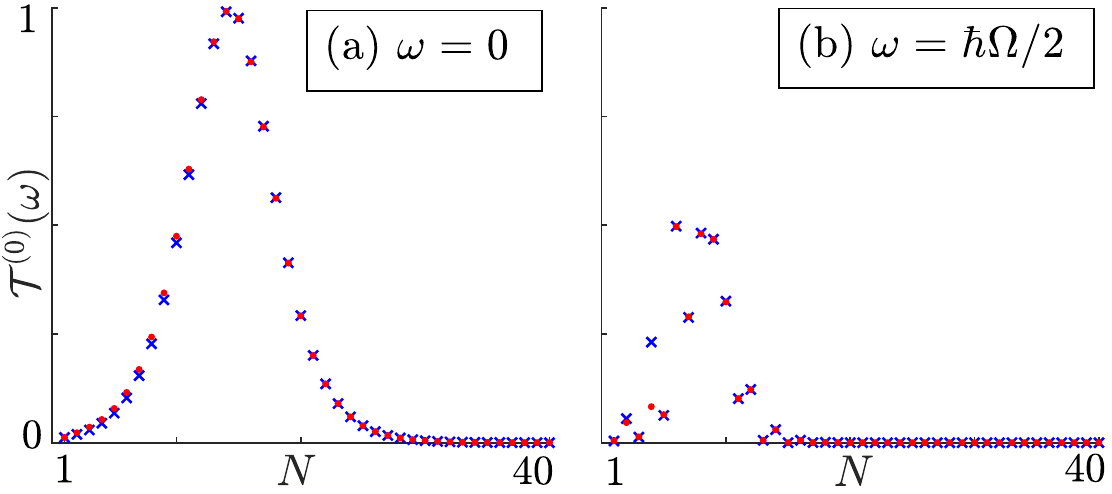}
    \caption{The photon-assisted transmission vs. number of dimers $N$ at (a) $\omega=0$; (b) $\omega=\hbar \Omega /2$. The results are obtained by using the numerical Green's function technique (blue) or by solving the problem within the projected space (red). The hopping amplitudes are set to $\gamma_{1,0} = 0.9$ and $\gamma_{2,0} = 1.1$, with driving $v(t) = 0.4 \cos(3t)$. The broadening functions are chosen as $\Gamma_{L,\, A 1} = \Gamma_{R,\, B N} = 0.1$.}
     \label{fig5}
\end{figure}

The transmission spectra $\mathcal{T}^{\, (0)}(\omega)$ for four different sets of model parameters corresponding to the four distinct topological phases of the harmonically driven dimer array is illustrated in Fig.~4. Here, and in the following figures, the data was obtained from a nonequilibrium Green's function approach, using Floquet-Sambe matrices numerically truncated to 7 rows and columns (for details, see Appendix~B). All four cases in Fig. 4 agree perfectly with the topological invariants calculated in Refs. \onlinecite{Lago}, \onlinecite{OBHJ2017}: By the bulk-boundary correspondence for Floquet topological insulators \cite{Asboth}, the presence of topological edge states $-$ signalled by single peaks in the transmission gap around $\omega = 0$ (``normal gap") and/or $\omega = \hbar \Omega/2$ (``anomalous gap") $-$ corresponds to nonzero values of the topological index $\nu_0$ (normal gap) and $\nu_{\pi}$ (anomalous gap). In this way transport characteristics fingerprint distinct topological phases in a periodically driven system.  

Similar to the undriven case, it is of obvious interest to understand how the transmission peaks change with the size of the FTI. For doing so we follow the idea used in the time-independent case (cf. Appendix~C), but now assisted by a diagonalization of the Hamiltonian in Floquet-Sambe space which allows us to numerically obtain the required parameter values, cf. Appendix~D. We apply this analysis to predict the scaling of the transmission peaks for the case of driven dimer arrays and then compare to the analogous result obtained from the somewhat cumbersome numerical Green's function method. The size-dependence of the transmission at $\omega = 0$ and $\omega=\hbar \Omega/2$, obtained using the two methods, is shown in~Fig.~5. The data points from the two approaches are in very good agreement with each other, also for small system sizes (except for $N=8$ when $\omega= \hbar \Omega/2$) where one does not expect such near-perfect agreement. The transmission at $\omega = 0$ behaves similarly to the time-independent case, Fig.~3. On the other hand, the behavior of the transmission at $\omega = \hbar \Omega/2$ is different. In~Fig.~6 the transmission at $\omega = \hbar \Omega/2$ is plotted as a function of not only array size but also of driving frequency~$\Omega$. It is interesting to note that the transmission oscillates as the frequency $\Omega$ is varied. The origin of such oscillations remains unknown to us and warrants a deeper investigation. We also find that the transmission peaks inside the anomalous gap do not exceed $1/2$ at any system size, Fig.~5, while in general a transmission amplitude is expected to be bounded by unity. Interestingly, this reduction of the transmission amplitude is due to the intrinsic structure of the symmetry-protected edges states at $\varepsilon = \hbar \Omega/2$: As described in detail in Appendix~D, the edge state components corresponding to the Floquet indices $m = -1$ and $m = 0$ have equal weights and from Eq. D9 only the latter component couples to the states incoming from the leads. In this way the transmission via the symmetry-protected states at $\varepsilon = \hbar \Omega/2$ is always reduced by a factor of $1/2$. The states at $\varepsilon = 0$ are, however, of a different structure, allowing the corresponding in-gap transmission peak to approach unity. This signifies an important difference between the two flavors of Floquet topological edge states. For a more complete and technical discussion refer to Appendix~D.

\begin{figure} \centering
    \includegraphics[width=8.5cm,angle=0]{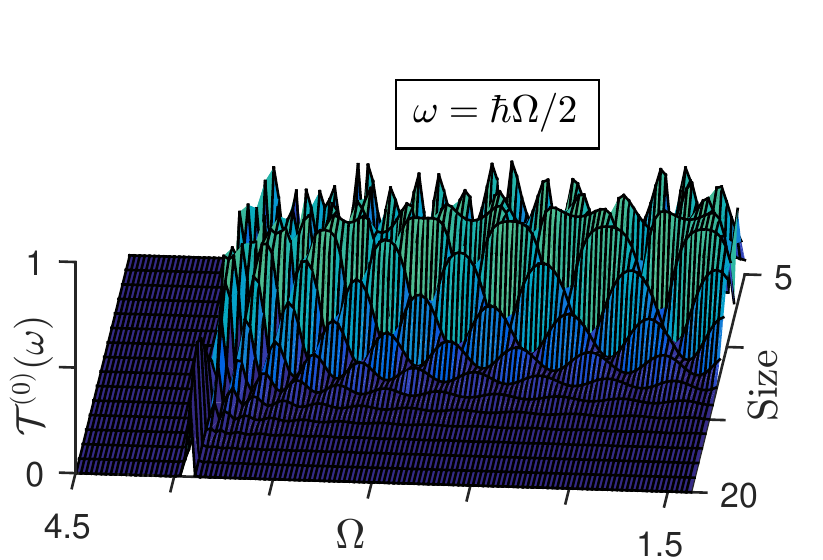}
    \caption{The transmission $\mathcal{T}^{(0)}$ at $\omega = \hbar \Omega/2$ vs. frequency $\Omega$ and number of dimers $N$. The driving is set to $v(t) = 0.4 \cos(\Omega t)$, the hopping amplitudes are $\gamma_{1,0} = 0.9$ and $\gamma_{2,0} = 1.1$, and the broadening functions $\Gamma_{L,\, A 1} = \Gamma_{R,\, B N} = 0.05$.}
     \label{fig6}
\end{figure}

\section{Symmetry Protection of The Edge States}

The edge states of topologically nontrivial insulators posses a variety of interesting properties distinguishing them from regular edge states, in particular, they are expected to be protected against symmetry-preserving local perturbations. For a 1D system in the thermodynamic limit the corresponding symmetry-protected states remain exactly in the middle of the bulk gap as long as the applied perturbation preserves certain symmetries and does not close the gap \cite{AsbothBook}. 
Here we propose that symmetry protection can be present to some extent (to be spelled out precisely below) also in a finite-size system, and can be seen in transport measurements performed on a perturbed finite-sized topological insulator.

\subsection{Symmetries of the SSH Model}

To set the stage, let us briefly recall that the static SSH Hamiltonian, Eq. (1) with $\gamma_1(t) = \gamma_{10}$ and $\gamma_2(t) = \gamma_{20}$, possesses a sublattice (chiral) symmetry \cite{AsbothBook}. This symmetry manifests the fact that there is no coupling between sites from the same sublattice. We define the chiral symmetry operator $\Gamma$ as the difference between the projectors onto the two sublattices $A$ and~$B$,
\begin{align}   
\begin{split}
\Gamma  =  \sum_{j} \left( |A,j  \rangle \langle A,j| - |B,j \rangle \langle B, j| \right),
\end{split}
\label{eq:SSH_chiral_symmetry}
\end{align}
from which one easily verifies that the SSH model indeed is chiral symmetric: $H = - \Gamma \, H \,\Gamma$. This symmetry forces the eigenmodes with opposite energies to come in pairs and therefore, in the thermodynamic limit, it requires that the zero-energy edge modes stay put at zero energy as long as the chiral symmetry is preserved and the bulk gap is open. It follows that a spatial disordering of hopping amplitudes $\gamma_1$ and $\gamma_2$ maintains the protection. On the contrary, the edge states are not robust against a disorder in the chemical potential~$\mu$ since the corresponding local operators $\sim | A/B, \, j \rangle \langle A/B, \, j |$ couple sites on the same sublattice and hence break chiral symmetry.

The periodically driven SSH model, Eq. (1) with driving chosen as in Sec.~II B ($\gamma_1(t) = \gamma_{1,0} - v(t)$ and $\gamma_2(t) = \gamma_{2,0} + v(t)$, with $v(t) \sim \cos(\Omega t)$), also exhibits chiral symmetry, now adapted to the Floquet formalism. The chiral symmetry within this theory is defined for the evolution operator $U (0, T)$ over one full period $T=2\pi/\Omega$ of the drive and is explicitly given by the relation $U (0, T) = \Gamma U^{-1} (0, T) \Gamma$. This symmetry remains unbroken as long as $H(t) = - \Gamma H( -t) \Gamma$ is fulfilled \cite{OBHJ2017}. One easily verifies that the chosen driving respects this condition. Moreover, in analogy to the time-independent case, the edge states are protected also against time-periodic perturbations which preserve the chiral symmetry. 
Examples of such perturbations include site-dependent disordering of the hopping amplitudes $\gamma_1$ and $\gamma_2$ that is even in
time (which, trivially, includes static perturbations), as well as perturbations from an added local chemical potential $\sim  | A/B, \, j \rangle \langle A/B, \,j |$
that is odd in time \cite{Asboth, OBHJ2017}. It follows that for these perturbations the edge modes have to stay put exactly at $\varepsilon=0$ or $\varepsilon=\hbar \Omega/2$ quasienergy.
It is important to note that the zero reference time is defined with respect to the bulk driving ($\sim \cos(\Omega t)$ in our case) and therefore the protection of the edge states crucially relies on the relative phase between the applied time-periodic perturbation and the bulk driving. 

\begin{figure} \centering
    \includegraphics[width=6.5cm,angle=0]{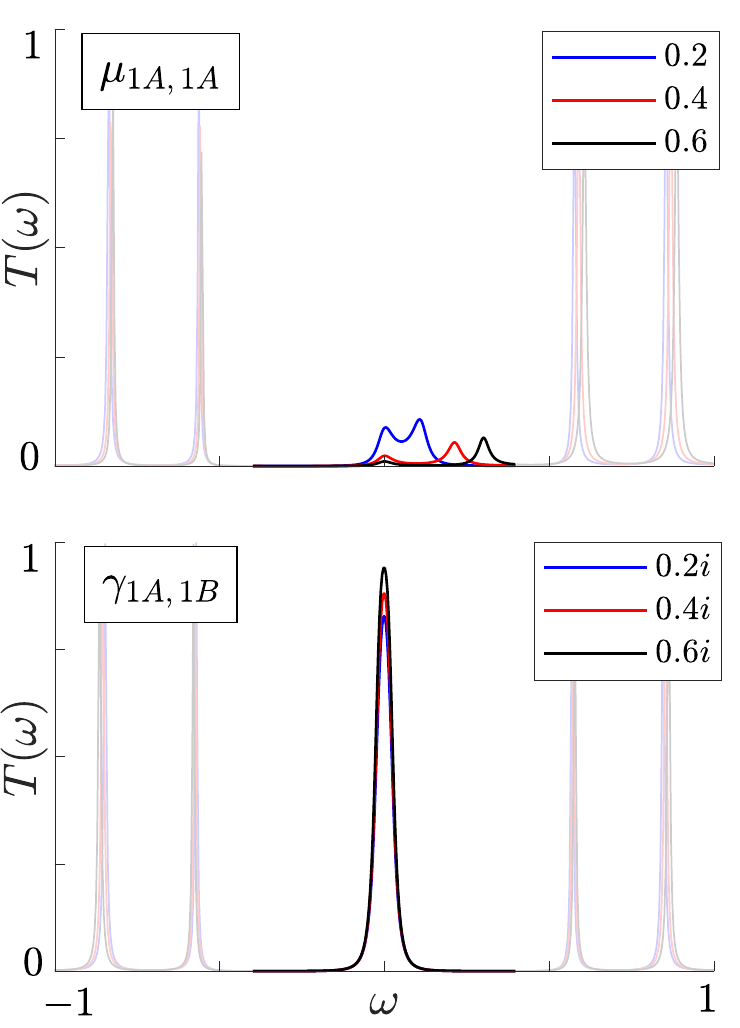}
    \caption{Transmission spectra of time-independent dimer arrays in the nontrivial topological phase, subject to static boundary perturbations in the chemical potential $\mu_{1A,\,1A}$ and intra-cell hopping amplitude $\gamma_{1A, 1B}$. The arrays consist of 9 dimers with hopping amplitudes $\gamma_{1} =0.8$, $\gamma_{2} = 1.2$. The transmission spectra are labeled (in color) by the disordering amplitudes used to perturb the edge states. The broadening functions are $\Gamma_{L, A 1} = \Gamma_{R,B N}= 0.1$.  The in-gap transmission spectra are distinguished by a stronger intensity in colors.}
     \label{fig7}
\end{figure}

One can also establish protection of the edge states in time-independent topological insulators against time-periodic perturbations. This is so because time-independent systems are trivially periodic in time. As a case in point, the undriven SSH model possesses a Floquet chiral symmetry, $U(t_0, t_0 + T) = \Gamma U^{-1} (t_0, t_0  + T) \Gamma$, for every reference time~$t_0$. This implies that the edge states stay protected as long as the chiral symmetry is unbroken for at least one choice of reference time $t_0$. Thus, the symmetry-protected edge modes will stay put at zero quasienergy for any harmonic disordering perturbation in $\gamma_1$, $\gamma_2$ or $\mu$ that acts at the boundary. Higher harmonics or disorders in a few parameters will in general break the symmetry and result in a quasienergy shift of an edge state away from zero or $\hbar\Omega/2$.

\subsection{Transport Properties: Time-Independent Topological Insulators Under Static Perturbations}

To pave the way, let us first consider the case of undriven 1D topological insulators and then extend our study to the periodically driven ones, the FTIs. As we saw in the previous section, edge states in a finite-sized topologically nontrivial system hybridize and by this build up transport channels. The hybridization destroys the protection since the boundary modes may overlap under perturbations, including symmetry-preserving ones, and for that reason may split in energy. The splitting will introduce an energy mismatch between distinct edge states and therefore threaten to kill the corresponding transmission peak. However, the edge modes overlap mainly in the middle of the insulator, with each mode quickly decaying when approaching the opposite side. In this way, any perturbation applied close enough to an edge has a very small effect on the state localized on the opposite edge of the system, and hence, its energy has to remain very close to zero. As for the effect of a perturbation on the state localized on the very same edge at which the perturbation acts, chiral symmetry ensures that its energy stays put at zero (or close to zero when hybridization with the opposite edge state is taken into account) if the perturbation respects chiral symmetry, otherwise the energy level will be shifted. This implies that under a symmetry-preserving boundary perturbation the transmission spectrum for a finite topological insulator will still contain a well-defined peak at zero energy. In contrast, if a symmetry-breaking perturbation is applied, the peak will disappear. This prediction can be formally justified by projecting the transport problem onto the space spanned by the edge states, similar to what we did in Sec.~II~A. It is then possible to analytically predict the behavior of the transmission peak and discuss to which extent it reflects the symmetry protection of the edge states. In short, one may prove that the leading-order correction to the zero-energy transmission amplitude vanishes exactly in the case of a symmetry-preserving boundary perturbation, leaving only the strongly suppressed corrections in play. The necessary derivation, being somewhat technical, can be found in Appendix~C.

Using our nonequilibrium Green's function approach, in Fig.~7 we present numerical results for a finite-size dimer array in the nontrivial topological phase, subject to a boundary perturbation in the chemical potential~$\mu_{1A, 1A}$, and in the intra-cell hopping amplitude~$\gamma_{1A, 1B}$ (with analogous results for a perturbation in the inter-cell hopping amplitude $\gamma_{1B, 2A}$). In perfect agreement with a state-space projection analysis and the discussion above, one observes a dramatic drop in the zero bias transmission under a boundary disorder in the chemical potential (symmetry-breaking) while the peak survives under boundary perturbations in hopping amplitudes (symmetry-preserving). As indicated in the panels of Fig.~7, the perturbations of the hopping amplitude are taken to be complex when doing the computations, in this way breaking particle-hole symmetry and leaving only the chiral symmetry to protect the edge states.  

Considering the symmetry-breaking perturbation, it is quite striking that the presence of the transmission peak can be suppressed by perturbing just one (or a few) sites, in this way switching the transport from conducting ($I_\text{dc}>0$) to insulating ($I_\text{dc} \simeq 0$). Let us also point out that if one applies a sufficiently strong symmetry-preserving boundary perturbation, the peak will first split into two and then gradually disappear. This is so because the edge state wavefunction carries some small but not negligible weight on the opposite side of the finite-sized topological insulator, and therefore the levels of the states may split under a large boundary perturbation, even if the symmetry is preserved. 

\begin{figure} \centering
    \includegraphics[width=6.5cm,angle=0]{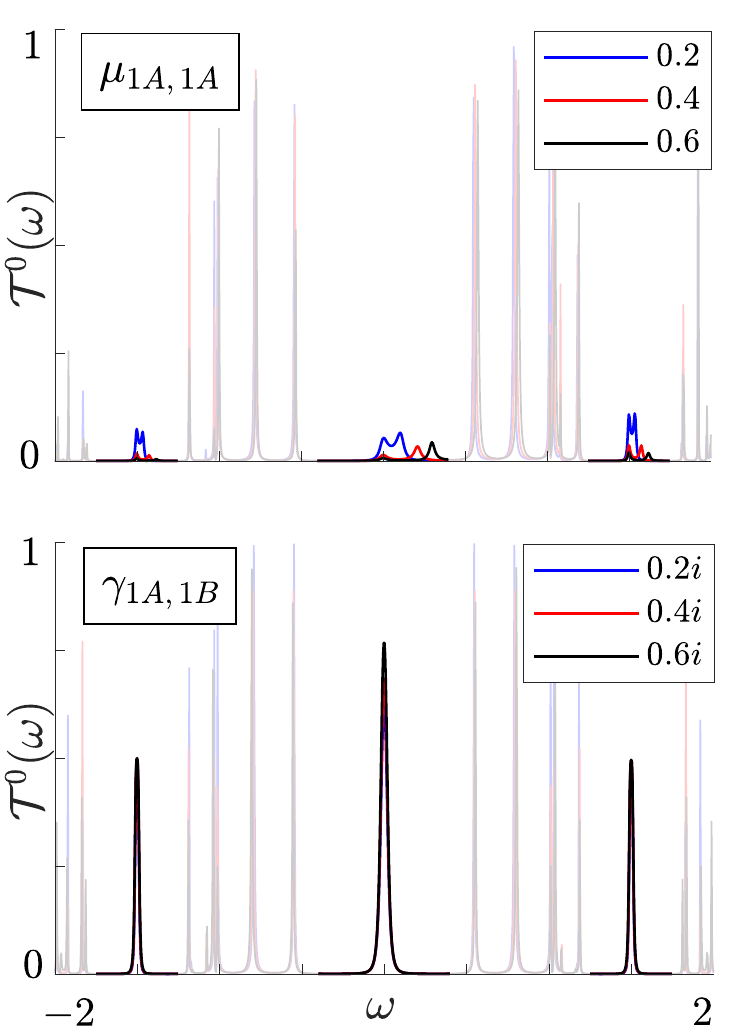}
    \caption{Transmission spectra of periodically driven dimer arrays in the topological phase with $\nu_0=1, \nu_{\pi}=1$, subject to time-independent boundary perturbations in the chemical potential $\mu_{1A,\,1A}$ and intra-cell hopping amplitude $\gamma_{1A, 1B}$. The arrays have 9 dimers and hopping amplitudes $\gamma_{1,0} =0.8$, $\gamma_{2,0} = 1.2$. The bulk driving is set to $v(t) = 0.4 \cos{(3 t)}$. The labels (in colors) of the different transmission spectra correspond to the amplitudes of the perturbation. The broadening functions are $\Gamma_{L, A 1} = \Gamma_{R,B N}= 0.1$. The in-gap spectra are plotted using higher color intensity.}
     \label{fig8}
\end{figure}
\subsection{Transport Properties: Floquet Topological Insulators Under Static Perturbations}

Following closely the discussion in Sec.~III B, we now extend the analysis to FTIs. The robustness of the transmission peaks at $\omega = 0$ and $\omega = \hbar \Omega/2$ against a boundary perturbation also for these systems requires that the perturbation respects chiral symmetry. Analogously to the time-independent case, this may be explained formally by viewing the problem within a projected state space. The idea is to analyze the time-periodic transport within Floquet-Sambe theory, by this facilitating a passage from the time-independent projected-state space analysis (cf. Appendix~C) to the present time-periodic case. It is then found that the leading-order correction to the mid-gap transmission also vanishes in this case provided that the chiral symmetry is maintained. The technical details can be found in Appendix~D. 

In Fig.~8  we present numerical results for periodically driven dimer arrays in the topological phase with $\nu_0=1, \, \nu_{\pi}=1$, subject to various static boundary perturbations in each of the parameters $\mu_{1A,\,1A}$ (chemical potential) and $\gamma_{1A, \, 1B}$ (intra-cell hopping amplitude). The behavior of the ``normal" and ``anomalous" midgap transmission peaks agree with the symmetry arguments: Both flavors of peaks (at $\omega = 0$ and $\omega = \hbar \Omega/2$ respectively) survive perturbations in the hopping amplitude while vanishing rapidly under perturbations in the chemical potential. As for the time-independent case in Sec.~III~B, the perturbations in the hopping amplitude were taken complex-valued in order to break the particle-hole symmetry which would otherwise have protected the edge states also when chiral symmetry is broken. 

\begin{figure} \centering
    \includegraphics[width=8.5cm,angle=0]{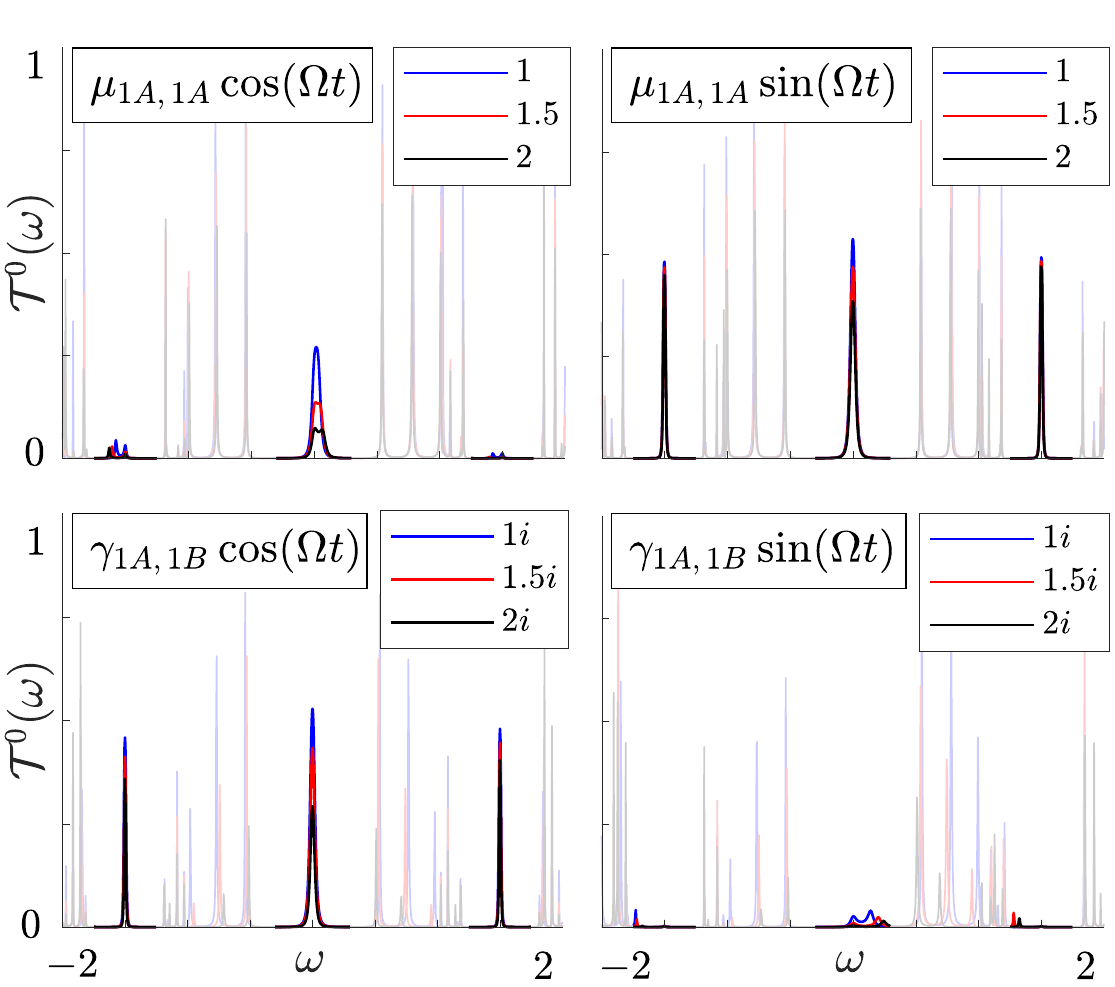}
        \caption{Transmission spectra of periodically driven dimer arrays in the topological phase with $\nu_0=1, \nu_{\pi}=1$, subject to time-periodic boundary perturbations in 
        $\mu_{1A,\,1A}$ (chemical potential) and $\gamma_{1A, 1B}$ (intra-cell hopping amplitude). The arrays are made up of 9 dimers with intra-cell and inter-cell hopping amplitudes $\gamma_{1,0} =0.8$ and $\gamma_{2,0} = 1.2$ respectively. The bulk driving is taken to be $v(t) = 0.4 \cos{(\Omega t)}$ with $\Omega = 3$. The periodic driving of the perturbations is either $\sim \cos(\Omega  t)$ or $\sim \sin(\Omega  t)$. The transmission spectra are labeled (in color) by the disordering amplitudes used to perturb the edge states. As before we consider the case with $\Gamma_{L, A 1} = \Gamma_{R,B N}= 0.1$. The in-gap spectra are plotted using higher color intensity.}
     \label{fig9}
\end{figure}

\begin{figure} \centering
    \includegraphics[width=6.5cm,angle=0]{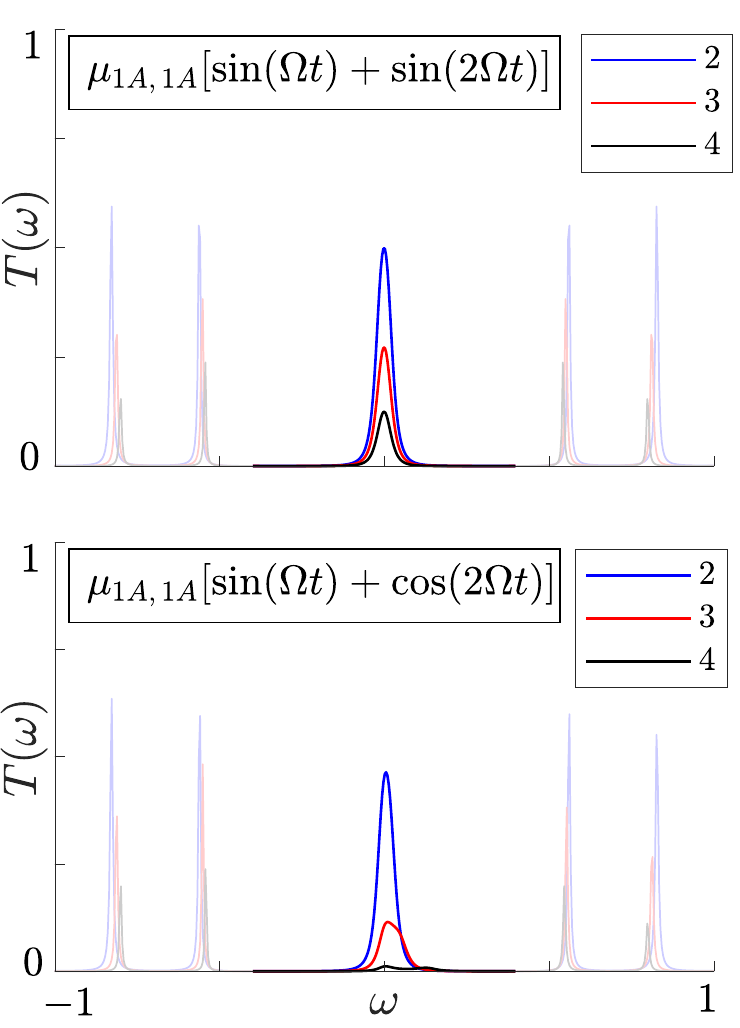}
    \caption{Transmission spectra corresponding to time-independent topologically nontrivial arrays of dimers perturbed by time-periodic disordering in $\mu_{1A, \, 1A}$ and $\gamma_{1A, \, 1B}$. The arrays are composed of 9 dimers with hopping parameters $\gamma_1 = 0.8$ and $\gamma_2 = 1.2$. The transmission spectra are labeled (in color) by the amplitudes of the applied perturbation. Each panel is labelled by the corresponding time-dependence of the perturbation. The broadening functions are set to $\Gamma_{L, A 1} = \Gamma_{R,B N}= 0.1$. The in-gap spectra are plotted using higher color intensity.}
     \label{fig10}
\end{figure}

\subsection{Transport Properties: Floquet Topological Insulators Under Time-Periodic Perturbations}

The issue of the stability of Floquet topological states under various types of time-dependent noise has attracted recent attention, both theoretically \cite{OBHJ2017,Rieder,Sieberer,Cadez} and experimentally \cite{Jorg,photonic1}. In the case of time-periodic {\em boundary perturbations}, the midgap edge states are known to be robust as long as the protecting symmetry is maintained \cite{OBHJ2017}. The time-dependence here grants an extra degree of flexibility compared to the perturbations discussed earlier and therefore these types of perturbations are probably of higher practical interest. In Fig.~9 we disturb arrays in a $\nu_0=1, \, \nu_{\pi}=1$ topological phase by a harmonic boundary perturbation in a single parameter (intra-cell hopping amplitude $\gamma_{1A, \, 1B}$ or chemical potential $\mu_{1A,\,1A}$) and vary the phase of the perturbation. Depending on the phase, the chiral symmetry may either be preserved or broken (cf. Sec.~III A). The obtained transmission spectra faithfully signal this fact: The peaks at $\omega = 0$ and $\omega =  \hbar \Omega/2$ only survive perturbations in hopping amplitudes (chemical potential) that are even (odd) in time. The most distinct transport signature of the symmetry protection is the behavior of the in-gap transmission spectra under perturbations in the same parameter and amplitude but with a phase mismatch in time. In Fig.~9 this qualitative difference is strikingly manifest.

In order to clearly see the anticipated effects we here had to increase the disordering amplitudes compared to the time-independent case. This is because the time-periodic eigenstates typically decay rapidly with the Floquet index~\cite{Rudner} and any time-periodic perturbation with zero mean couples only state components corresponding to different Floquet indices. Interestingly, we also notice that under time-periodic perturbations the transmission peaks in the anomalous gap (at $\omega =  \hbar \Omega/2$) are more sensitive to a breaking of the chiral symmetry than the peaks in the normal gap (at $\omega = 0$), cf. Fig.~9. This is a generic feature originating from the structural difference of the Floquet topological edge states at $\varepsilon = 0$ and $\varepsilon = \hbar \Omega/2$. The chiral symmetry obligates Floquet components of the symmetry-protected edge states to come in pairs with equal weights (Appendix D): the Floquet component $m$ is paired with the component $-m$ $(-m-1)$ in the case of a symmetry-protected edge state at $\varepsilon = 0$ $(\varepsilon = \hbar \Omega/2)$. It follows that any time-periodic perturbation with zero mean, i.e. a perturbation which couples only components corresponding to different Floquet indices, will have generically larger effect on the transmission via the states at $\varepsilon = \hbar \Omega/2$ than at $\varepsilon = 0$, see Appendix D. For example, let us take a perturbation $p(t) \sim \cos(\Omega t + \phi)$. It couples Floquet component $m$  with components $m \pm 1$. The coupled components $m = 0$ and $m - 1 = -1$ are {\it both} large for the symmetry-protected edge states at $\varepsilon = \hbar \Omega/2$, however, only component $m = 0$ is large for the states at  $\varepsilon = 0$. This means that the overall effect of $p(t)$ on the edge states and transmission at zero will be much smaller than at $\hbar \Omega/2$. Note that such a structural difference between the edge states at $\varepsilon = 0$ and $\varepsilon =  \hbar \Omega/2$ is reflected in the transmission spectra only under \textit{time-periodic} symmetry-breaking perturbations, and one should not expect any similar effect under influence of \textit{time-independent} perturbations, in agreement with the results in Fig.~8.


\subsection{Transport Properties: Time-Independent Topological Insulators Under Floquet Perturbations}

It is also interesting to look at undriven 1D topological insulators subject to time-periodic boundary perturbations. In this case the edge states should also exhibit Floquet symmetry protection \cite{OBHJ2017}, and one expects such setups to be easier to implement in experiments since they require less control.

The midgap transmission peaks of the topologically nontrivial undriven arrays are expected to be only slightly affected by the same time-periodic perturbations as for the driven arrays considered earlier (cf. Fig.~9). This is because for these cases there is always a reference time $t_0$ for which the Floquet chiral symmetry is preserved \cite{OBHJ2017}.
However, adding a higher out-of-phase harmonic to the perturbation breaks the chiral symmetry for  {\em all} reference times~$t_0$ and the transmission peak is expected to disappear much faster than if the perturbation is symmetry-preserving for at least {\em one} reference time~$t_0$. This behaviour is confirmed in Fig.~10. To see the difference in the decay rates we had to go to very large disordering amplitudes since the quasienergy shift induced by such symmetry-breaking perturbation is found to be extremely small \cite{OBHJ2017}. In fact, the quasienergy splitting may be even smaller than the quasienergy difference due to the finite-size effect. Therefore, in both symmetry-preserving and symmetry-breaking cases the transmission goes down first at roughly the same rate. At larger perturbation amplitudes, however, the different behaviors of the mid-gap transmissions become distinguishable.

 \section{Summary and Outlook}

We have studied a setup for obtaining transport signatures of the robustness of topological edge states in a 1D finite-sized Floquet topological insulator in the presence of local boundary perturbations. Using a Su-Schrieffer-Heeger model of a periodically driven dimer array as minimal model, and exploiting the Floquet-Sambe Green's function approach for Floquet transport \cite{OB2018}, we
have obtained the transmission spectra encoding the zero-temperature linear conductances for different types of perturbations. The results, 
with time-independent and periodically driven local boundary perturbations which either respect or violate the protecting chiral symmetry $-$ boost our proposal for observing the symmetry protection of Floquet topological edge states: A perturbation which respects (violates) the symmetry correlates perfectly with the presence (absence) of a distinct mid-gap peak in the transmission spectrum, both for the normal and ``anomalous" quasienergy gap. The stabilities of the peaks do not differ significantly between the two gaps, although the transmission in the anomalous gap appears more fragile against symmetry-breaking perturbations. Recent realizations of the SSH model in the solid state\cite{Drost, Belopolski}, with cold atoms \cite{coldatoms1,Zheng,coldatoms2}, and in photonic crystals \cite{Zhou,photonic2,photonic1}, suggest that an experimental test, while being probably challenging to implement in practice, could become possible in the near future. The recent breakthroughs in measuring transport using two-terminal setups with cold atoms may here hold particular promise 
\cite{Krinner1,Krinner2}.

Although our study has focused on 1D Floquet topological insulators protected by chiral symmetry, the approach presented is quite general and should be possible to extend to systems in higher dimensions. In particular, it would be interesting to see how it plays out for time-reversal symmetric topological insulators in 2D and 3D where the periodic drive breaks the continuous time-reversal symmetry down to a discrete symmetry.

\section*{ACKNOWLEDGMENTS}

We thank Sebastian Eggert and Teemu Ojanen for valuable communications. This work was supported by the Swedish Research Council through Grant No. 621-2014-5972. \\ 

\appendix

\section{Details on the Green's function formalism: Time-independent}

Here we provide details on the non-equilibrium Green's function theory used in this work. Transport properties can be extracted from the (retarded) Green's function $G_d(\omega)$ defined as   
\begin{align}   
\begin{split}
G_d(\omega) = \left[\omega^{+}  - H_d - \Sigma_L(\omega) - \Sigma_R(\omega) \right]^{-1},
\end{split}
\label{eq:Greens_function}
\end{align}
where $H_d$ is the Hamiltonian of the central (``device") system, $\Sigma_{\nu}(\omega)$ ($\nu = L,R$) are self-energies representing the leads, and $\omega^{+} = \omega + i \eta$ with infinitesimal $\eta>0$. 

Within the Landauer-B\"{u}ttiker formalism\cite{Datta}, the transmission spectrum $T(\omega)$ defined in Eq.~(\ref{eq:dc_current}) can be obtained from the relation 
\begin{align}   
\begin{split}
T(\omega) = \text{Tr} [G_d^\dagger \Gamma_L G_d \Gamma_R ](\omega),
\end{split}
\label{eq:Transmission}
\end{align}
with broadening functions $\Gamma_{\nu}(\omega) = i(\Sigma_{\nu}(\omega) - \Sigma^\dagger_{\nu}(\omega))$. We shall simplify the problem by applying the so-called wide-band limit approximation \cite{WBL}. This assumes that the density of states in the leads does not vary much near the Fermi energy and therefore can be taken constant. It follows that $\Sigma_{\nu}(\omega)$ can be considered independent of~$\omega$. Also, within this approximation we do not take into account shifts of the Green's function poles: $\text{Re}(\Sigma_{\nu}(\omega))$ is set to zero and the self-energies take the simple form $\Sigma_{\nu} = - i \, \Gamma_{\nu}/2$, where $\Gamma_{\nu}$ are energy-independent real symmetric matrices. The wide-band limit approximation well represents metallic leads \cite{WBL} and simplifies computations significantly.

In general, the self-energies $\Sigma_{\nu}(\omega)$ describe tunneling between the leads and the transport region. For our case study we assume that the coupling is only between the leads and the first/last sites of the array. Therefore, the matrices $\Sigma_{\nu}(\omega)$ are zero everywhere except for the first ($\nu = L$) or last ($\nu = R$) diagonal entries. Thus, from~Eq.~(\ref{eq:Transmission}), the transmission is given by 
\begin{align}   
\begin{split}
T(\omega) =  \Gamma_{L,\, A 1} \Gamma_{R,\, B N} | G_{(A, 1 | B, N)}(\omega)|^2,
\end{split}
\label{eq:Transmission_reduced}
\end{align}
 where $\Gamma_{L,\, A 1}$ and $\Gamma_{R,\, B N}$ are the only non-zero elements of the matrices $\Gamma_L$ and $\Gamma_R$ respectively, and $G_{ (A, 1 | B, N)}(\omega)$ is the corresponding matrix entry of the Green's function $G_d(\omega)$.  
 
A similar expression should be also applicable to any 1D topological insulator attached to two leads at the edges, provided that the formula is properly adjusted to include all nonzero coupling elements in the broadening functions $\Gamma_{\nu}(\omega)\, (\nu=L,R)$. In this general case, the sublattices $A$ and $B$ are defined using the sublattice projection operators $P_A = (\mathbb{1} + \Gamma)/2$ and $P_B = (\mathbb{1} - \Gamma)/2$, where $\mathbb{1}$ is the identity operator and $\Gamma$ is the chiral operator which performs the transformation under which the Hamiltonian is invariant.

\section{Details on the Green's function formalism: Periodically-driven}

Transport through periodically driven systems can be addressed in a similar way as for the undriven case, see Ref. \onlinecite{OB2018}. The idea is to go to the so-called Floquet-Sambe space \cite{Sambe}, transforming the periodically driven problem into an equivalent time-independent one. Within such a representation the time-periodic Hamiltonian $H(t)$ turns into a Floquet-Sambe Hamiltonian $\mathcal{H}$, defined by 
\begin{align}
\begin{split}
\llangle i, m| \mathcal{H} |j, m^\prime \rrangle &= \int_0^T dt   \langle i|  e^{-i m\Omega t} \left[H(t) - i\hbar \partial_t \right]e^{i m^\prime\Omega t}|j \rangle \\
&=  \langle i|\left[ H^{(m^\prime - m)} + m \hbar \Omega \mathbb{1} \delta_{m, m^\prime}  \right] |j \rangle,
\end{split}
\label{eq:Sambe1}
\end{align}
where $\langle i| H^{(n)} |j \rangle = T^{-1}\int^T_0 \, dt \, e^{i n \Omega t} \langle i| H(t) |j \rangle$ and $\mathbb{1}$ is the identity operator. The notation $| \, ... \, \rrangle$ here represents a time-periodic state with period $T$, in particular $|i, m \rrangle = e^{i m \Omega t} |i\rangle$. The inner product between the time-periodic states is obtained by time-averaging the conventional inner product over one period. Explicitly, the Floquet-Sambe Hamiltonian reads as follows:
\begin{align}
\begin{split}
\mathcal{H} =  \begin{pmatrix}
 \ddots & \vdots & \vdots  & \vdots  & \udots\\
  ...  & H^{(0)} - \hbar\Omega \mathbb{1} & H^{(1)} & H^{(2)} & ...  \\  
 ... & H^{(-1)} & H^{(0)} & H^{(1)} & ... \\ 
...  & H^{(-2)} & H^{(-1)} & H^{(0)} + \hbar\Omega \mathbb{1} & ...  \\ 
  \udots  & \vdots  & \vdots  & \vdots  & \ddots\\
\end{pmatrix}.
\end{split}
\label{eq:Sambe2}
\end{align}

Here we consider a finite-sized central system driven by time-periodic external gates and subject to a time-independent voltage (from the bias between the electrochemical potentials of the two leads contacted to the system).  Within this scenario the dc current may be obtained through Eq.~(\ref{eq:current_total_FS}) with photon-assisted transmissions $\mathcal{T}^{\, (0)}_{\nu, \nu^\prime}$ ($\nu, \nu^\prime = L,R$) defined by \cite{OB2018}
\begin{align}   
\begin{split}
\mathcal{T}^{\, (0)}_{\nu, \nu^\prime} (\omega) = \text{Tr} \left[\mathcal{G}^\dagger_d \varGamma_{\nu^\prime}  \mathcal{G}^{\phantom\dagger}_d  \varGamma^{(0)}_{\nu} \right] (\omega).
\end{split}
\label{eq:transmission_floquet}
\end{align}
We have here used the notation from Ref. \onlinecite{OB2018}: $ \mathcal{G}_{d} = \left[\omega^+ I - \mathcal{H}_d - \mathcal{E}_L - \mathcal{E}_R  \right]^{-1} $ is the (retarded) Floquet-Sambe Green's function corresponding to a time-periodic Hamiltonian $H_d(t)$ of the driven central system, \text{$\varGamma_\nu = i(\mathcal{E}_\nu - \mathcal{E}^\dagger_{\nu})$} are Floquet-Sambe broadening functions with $\mathcal{E}_\nu$ Floquet-Sambe self-energies. Within the considered class of systems the Floquet-Sambe self-energies are given by
\begin{align}
\begin{split}
\mathcal{E}_\nu =  \begin{pmatrix}
 \ddots & \vdots & \vdots  & \vdots  & \udots\\
  ...  & \Sigma_\nu(\omega+\hbar\Omega) & 0 & 0 & ...  \\  
 ... & 0& \Sigma_\nu(\omega) & 0 & ... \\ 
...  & 0 & 0 & \Sigma_\nu(\omega - \hbar\Omega) & ...  \\ 
  \udots  & \vdots  & \vdots  & \vdots  & \ddots\\
\end{pmatrix},
\end{split}
\label{eq:Sambe_lead}
\end{align}
with conventional time-independent self-energies $\Sigma_\nu(\omega)$ $\text{($\nu = L, R$)}$. Finally, $\varGamma^{(0)}_{\nu^\prime}$ in Eq.~(\ref{eq:transmission_floquet}) represents the Floquet-Sambe zero matrix with the $(m, m^\prime) = (0,0)$ block being replaced by the time-independent broadening function $\Gamma_{\nu^\prime}$. For more details we refer the reader to Ref. \onlinecite{OB2018}.

In analogy to the time-independent case we simplify the description by employing the wide-band limit approximation, now extended to the Floquet formalism: It is assumed that $\mathcal{E}_{\nu}(\omega)$ is $\omega$-independent and is equal to $- i \, \varGamma_{\nu}/2$. Clearly, this assumption is valid as long as $\Sigma_{\nu}(\omega)$ does not vary much near the energy window $\omega \in [E_{F} - m_\text{cut} \hbar \Omega, E_{F} + m_\text{cut} \hbar \Omega]$ with equilibrium Fermi energy $E_{F}$ and cutoff $m_\text{cut}$ of the Floquet-Sambe space. Usually, $m_\text{cut}$ (with
$(2 m_\text{cut}+1)^2$ counting the number of blocks in the Floquet-Sambe matrix in Eq. (\ref{eq:Sambe2}))  is taken as a small number \cite{OB2018, Rudner}, yielding an accurate  approximation for most metallic leads. One should here note that a steady state of a periodically driven system attached to leads with a wide bandwidth may be affected by radiative recombination and electron-phonon scattering, present in a real experiment but neglected here. ``Energy-filters" suppressing multi-photon tunnelings will in general significantly reduce these effects \cite{Seetharam}. Our results, both numerical and analytical, can be adapted to include such filtered transport and we have verified that the results obtained within the wide-band limit approximation do not change much in this case. 

For our case study, using the harmonically driven SSH model in Eq. (\ref{eq:SSH_original}), we consider symmetric couplings between the leads and the first/last sites in the array, and thus arrive 
at the following expressions for the total photon-assisted transmissions:
\begin{align}   
\begin{split}
\mathcal{T}_{LR}^{\, (0)}(\omega) = \text{Tr} [\mathcal{G}^\dagger_{ (A, 1 | B, N)}\varGamma_{R,\, B N} \mathcal{G}_{ (B, N | A, 1)} \varGamma^{(0)}_{L,\, A 1}],
\end{split}
\label{eq:Transmission_reduced_Floquet12}
\end{align}
\begin{align}   
\begin{split}
\mathcal{T}_{RL}^{\, (0)}(\omega) = \text{Tr} [\mathcal{G}^\dagger_{ (B, N| A, 1)}\varGamma_{L,\, A 1} \mathcal{G}_{ (A, 1 | B, N)} \varGamma^{(0)}_{R,\, B N}].
\end{split}
\label{eq:Transmission_reduced_Floquet21}
\end{align}
Here the trace is taken over the Floquet-Sambe diagonal blocks. The indices on $\mathcal{G}$ and $\varGamma$ are the same as the indices in Eq. (\ref{eq:Transmission_reduced}). Note that in general $\mathcal{T}^{\, (0)}_{LR} \neq \mathcal{T}^{\, (0)}_{RL}$, even when the left and right broadening functions are the same, $\varGamma_{L,\, A 1} = \varGamma_{R,\, B N}$. One expects similar expressions to be applicable for calculating the transmission spectrum through any 1D FTI protected by chiral symmetry.

\section{The projected space representation: Time-independent}

The transport properties for applied biases much smaller than the width of the bulk band gap can be efficiently addressed by projecting the transport problem onto the states closest to zero energy, i.e., the edge states. At finite sizes of a topologically nontrivial system protected by a chiral symmetry (cf. Sec. III.~A), the edge states are modified by the confinement and split in energy. Moreover, the chiral symmetry restricts them to be of the form $|\psi^0_+ \rangle = (|\psi^0_A \rangle + |\psi^0_B \rangle)/\sqrt{2}$ with energy $E^0$ and $|\psi^0_- \rangle = (|\psi^0_A \rangle - |\psi^0_B \rangle)/\sqrt{2}$ with energy $-E^0$. Here the states $|\psi^0_A \rangle$ and $|\psi^0_B \rangle$ represent the projections to the sublattices $A$ and $B$ defined using the projection operators $P_A$ and $P_B$, cf. Appendix~A. Now, we assume that the high-energy bulk states do not contribute to the transport properties at biases much smaller than the width of the bulk band gap. We may then treat the problem in the projected space $\mathcal{S}^0$ spanned by $|\psi^0_\pm \rangle$. While the analysis to follow can be generalized to the case of multiple symmetry-protected states at the boundaries, we here assume that there are only two edge states present. By the structure of these states, $\mathcal{S}^0$ is also spanned by $|\psi^0_A \rangle$ and $|\psi^0_B \rangle$. Note that in the thermodynamic limit $N \rightarrow \infty$ the states  $|\psi^0_A\rangle$ and  $|\psi^0_B\rangle$ are equal to the symmetry-protected edge states because each of them lives on a different sublattice. Without loss of generality, $|\psi^0_A\rangle$ and $|\psi^0_B\rangle$ are assumed to correspond to the left and right boundaries respectively. We may also include perturbations in the discussion: If the resulting level splitting is small in comparison to the bulk gap, then we may still perform all calculations in $\mathcal{S}^0$ and get essentially exact results. 
\\

\textit{Unperturbed system:} A Hamiltonian $H_d$ of a generic 1D chiral-symmetric topological insulator projected onto the space $\mathcal{S}^0$ is given by
\begin{align}   
\begin{split}
H_d^{0} = \begin{bmatrix}
   \langle\psi^0_A |H_d|\psi^0_A \rangle      & \langle\psi^0_A |H_d|\psi^0_B \rangle  \\
    \langle\psi^0_B |H_d|\psi^0_A \rangle & \langle\psi^0_B |H_d|\psi^0_B \rangle  \\
    \end{bmatrix}
    = 
    \begin{bmatrix} 0     & \tau^0  \\
    (\tau^0)^\dagger & 0 
\end{bmatrix},
\end{split}
\label{eq:projected_H}
\end{align}
where $ \tau^0 = \langle\psi^0_A |H_d|\psi^0_B \rangle$. The diagonal terms are zero because $H_d$ couples only sites from different sublattices. Clearly, $H_d^{0}$ has eigenstates $|\psi^0_\pm \rangle$ with energies $\pm E^0 = \pm |\tau^0|$.

The (retarded) Green's function in $\mathcal{S}^0$ is given~by 
\begin{align}   
\begin{split}
G^{0}_d (\omega) = (\omega^+ - H_d^{0}  - \Sigma_L^0 -\Sigma_R^0)^{-1} =
    \begin{bmatrix} \omega^+ - \Sigma^{0}_L     & \tau^0  \\
    (\tau^0)^\dagger & \omega^+ - \Sigma^{0}_R
\end{bmatrix}^{-1}\!,
\end{split}
\label{eq:projected_G}
\end{align}
where $\Sigma^{0}_{L/R}$ are self-energies in $\mathcal{S}^{0}$. Here the self-energy $\Sigma_{L}$ ($\Sigma_{R}$) that gets projected onto $\mathcal{S}^{0}$ couples state $|\psi_A \rangle$ ($|\psi_B \rangle$) only to lead $L (R)$. In the most general case, other self-energy projection terms are also present, however, each of them is proportional to either $\Sigma_{L}|\psi_B \rangle$ or $\Sigma_{R}|\psi_A \rangle$ and therefore is strongly suppressed.

Now, by projecting Eq. (\ref{eq:Transmission_reduced}) onto $\mathcal{S}^{0}$, one obtains, in obvious notation:
\begin{align}
\begin{split}
T^0(\omega) &= \Gamma^0_L \Gamma^0_R |G^0_{12} (\omega)|^2 \\ &= \frac{\Gamma^0_L \Gamma^0_R |\tau^0|^2}{|(\omega^+ - \Sigma^0_L)(\omega^+ - \Sigma^0_R) - |\tau^0|^2|^2}.
\end{split}
\end{align}
We are interested in the scaling of the transmission at zero bias:
\begin{align}   
\begin{split}
T^0(0) = \frac{\Gamma^0_L \Gamma^0_R (E^0)^2}{|\Gamma^0_L \Gamma^0_R / 4 +  (E^0)^2|^2},
\end{split}
\label{eq:Transmission_proj}
\end{align}
where $\Sigma^0_L =  -i \, \Gamma^0_L/2$  and $\Sigma^0_R = -i \, \Gamma^0_R/2$ within the wide-band limit approximation. The projected self-energies 
$\Sigma^0_L = \sum_j \Sigma_{L,Aj} |\langle j, A |\psi^0_A \rangle|^2$ and $\Sigma^0_R = \sum_j \Sigma_{R, Bj} |\langle j, B |\psi^0_B \rangle|^2$ are proportional to the wavefunction amplitudes at the two edges of the topological insulator where the coupling to the leads is non-zero. The wavefunctions $|\psi^0_{A/B} \rangle$ and energy $E^0$ needed for computing $\Sigma_\nu^0$ and $|\tau^0|^2$ may be retrieved from a numerical diagonalization of the Hamiltonian $H_d$.

For our case study, a topologically nontrivial array of dimers described by the SSH Hamiltonian in Eq.~(1), one may develop a fully analytic prediction. We exploit the fact that it is straightforward to analytically find the two zero-energy edge modes of~$H_d$ in the thermodynamic limit. These modes may then be used to approximate the states $|\psi^0_{A/B} \rangle$, simply by truncating them at the applicable size of the array. It follows that, using the fact that lead $L (R)$ couples only to the first (last) site of the array, the energy $E^0$ is obtained from
\begin{align}   
\begin{split}
(E^0)^2 = \left|\gamma_1 (\gamma_1/\gamma_2)^{N-1}  |\langle 1, A |\psi^0_A \rangle|^2 \right| ^2,
\end{split}
\label{eq:wavefunction}
\end{align}
where
\begin{align}   
\begin{split}
|\langle 1, A |\psi^0_A \rangle|^2 =  |\langle N, B |\psi^0_B \rangle|^2 = \frac{1 - |\gamma_1/\gamma_2|^{2} }{1 - |\gamma_1/\gamma_2|^{2N}},
\end{split}
\label{eq:E}
\end{align}
and where $\gamma_{1/2}$ are the hopping amplitudes defined in Eq.~(1), with $N$ the number of dimers in the array. 
\\

\textit{Perturbed system:} We now establish the resilience of the projected transmission spectrum to a general symmetry-preserving local boundary perturbation represented by an operator $p$, acting at, say, the left boundary of the topological insulator. In the presence of $p$, the projected Green's function takes the form 
\begin{align}   
\begin{split}
G_d^0 (\omega) = \begin{bmatrix} \omega^+ - \Sigma^0_L + \delta h^0_L     & (\tau^0 + \delta  \tau^0)  \\
    (\tau^0 + \delta  \tau^0)^\dagger & \omega^+ - \Sigma^0_R + \delta h^0_R
\end{bmatrix}^{-1},
\end{split}
\label{eq:projected_0_perturbed}
\end{align}
where $\delta h^0_{L/R} =  \langle \psi^0_{A/B} |\, p \,|\psi^0_{A/B} \rangle$ and $\delta  \tau^0 = \langle \psi^0_A |\,p \,|\psi^0_B \rangle$. We now employ the fact that $p$ is a local perturbation acting at the left edge and $|\psi^0_B\rangle$ is localized at the right edge, which means that $\delta h^0_R$ is strongly suppressed. The term~$\delta  \tau^0$ is also small because of the same reason, even though it is expected to be larger than $\delta h^0_R$. The projected transmission can be affected in a significant way only by the term $\delta h^0_L$ which consists of an overlap between $p |\psi^0_A\rangle$ and $|\psi^0_A\rangle$. However, for a chirally invariant perturbation $p$ (satisfying the relation  $\Gamma \, p \, \Gamma = - p$), the correction $\delta h^0_L$ vanishes identically. This is because $\delta h^0_L = 
 \langle \psi^0_{A} |\,p  \,|\psi^0_{A} (t) \rangle = -\delta h^0_L = 0$, using the property that $|\psi^0_A \rangle = \Gamma |\psi^0_A  \rangle$. 
In this way the transmission peak reflects a resilience of the edge states against symmetry-preserving perturbations, as was anticipated in Sec. III. 

Now we return to our case study of a dimer array connected to the leads, and first consider a symmetry-breaking perturbation in the chemical potential $\mu_{1A, 1A}$ added to the first site in the array. For this case the projected Green's function reads as follows:
\begin{align}   
\begin{split}
G_d^0 (\omega) = \begin{bmatrix} \omega^+ -  \mu^0 - \Sigma^0_L     & \tau^0  \\
    (\tau^0)^\dagger & \omega^+ - \Sigma^0_R
\end{bmatrix}^{-1},
\end{split}
\label{eq:projected_H_mu}
\end{align}
with transmission 
\begin{align}   
\begin{split}
T^0_\mu(\omega)  &= \Gamma^0_L \Gamma^0_R |G^0_{12} (\omega)|^2 \\
& = \frac{\Gamma^0_L \Gamma^0_R |\tau^0|^2}{|(\omega^+- \Sigma^0_L - \mu^0)(\omega^+ - \Sigma^0_R) - |\tau^0|^2|^2},
\end{split}
\end{align}
where $\mu^0$ is the projected amplitude of the perturbation, $\mu^0 = |\langle 1, A |\psi^0_A \rangle|^2 \mu_{1A, 1A}$. Therefore, within the wide-band limit approximation we get 
\begin{equation}   
T^0_\mu(0) =\frac{\Gamma_L^0 \Gamma_R^0 (E^0)^2}{ |\Gamma_L^0 \Gamma_R^0 / 4 +  (E^0)^2|^2 + (\mu^0)^2 (\Gamma_R^0)^2 / 4},
\end{equation}
which can be expressed as
\begin{equation}
[T^0_\mu(0)]^{-1}  = [T^0(0)]^{-1} + \frac{\Gamma_R^0(\mu^0)^2}{4\Gamma_L^0(E^0)^2},
\end{equation}
with $T^0(0)$ given in Eq. (\ref{eq:Transmission_proj}). As expected, the transmission is seen to drop with~$\mu_{1A, 1A}$.

Next, we add a symmetry-preserving perturbation in the hopping amplitude between the first and second sites of the array, call it $\gamma_{1A, 1B}$. As a result, the corresponding projected Green's function gets expressed as 
\begin{align}   
\begin{split}
G^0_\gamma (\omega) = \begin{bmatrix} \omega^+  - \Sigma^0_L     & (\tau^0 + \delta \tau^0) \\
    (\tau^0 + \delta \tau^0)^\dagger & \omega^+ - \Sigma^0_R
\end{bmatrix}^{-1}
\end{split}
\label{eq:projected_H_hopping}
\end{align}
where  $\delta \tau^0$ is the projected perturbation, $\delta \tau^0 =  \langle \psi^0_A |1, A \rangle \langle 1, B |\psi^0_B \rangle \gamma_{1A, 1B}$. As was discussed earlier in a more general setting, this term is extremely small and vanishes completely in the thermodynamic limit since, in the topological phase, $\langle 1, B |\psi^0_B \rangle \sim (\gamma_1/\gamma_2)^N \ll 1$. Thus, one does not expect a significant change in the transmission unless the perturbation is extremely large, in case employing the projected space $S^0$ becomes inappropriate.

\section{The projected space representation: Periodically-driven}

In periodically driven chiral systems the driving may open an additional gap, a so-called dynamical (or ``anomalous") gap, which may also host symmetry-protected edge states \cite{Rudner,Carpentier}. The chiral symmetry in the periodically driven case, $\Gamma H_d(t) \Gamma = - H_d (-t)$ with $\Gamma$ the chiral symmetry operator, obligates the Floquet steady states \cite{Sambe} to come in pairs:  $|\psi (t) \rangle$ is a steady state with quasienergy $\varepsilon$ if and only if  $\Gamma|\psi (-t) \rangle$ is also a steady state but with quasienergy~$-\varepsilon$. 

Here we follow a very similar strategy to that for the time-independent case: We project the transport problem onto the space spanned by a pair of midgap edge states. There are two different quasienergy bulk gaps in the periodically driven case and we will consider them separately in the two subsections to follow. Let us note in passing that the more general case with multiple symmetry-protected edge states can be treated within the same formalism, here applied to the case when there is only a single pair of edge states at each quasienergy gap.  

\subsection{The pair of midgap edge states near zero quasienergy}

The chiral symmetry requires Floquet edge states to form pairs of the form $|\psi^0_{I} (t) \rangle$ and $|\psi^0_{II} (t) \rangle = \Gamma|\psi^0_{I} (-t) \rangle$ with quasienergies 
$\varepsilon^0$ and $-\varepsilon^0$ respectively. Accordingly, the space spanned by these modes is also spanned by the two states $|\psi^0_\pm (t) \rangle = (|\psi^0_{I} (t) \rangle \pm |\psi^0_{II} (t) \rangle)/\sqrt{2}$. These states satisfy $|\psi^0_\pm (t) \rangle = \pm \Gamma|\psi^0_\pm (-t) \rangle$, or equivalently $|\psi^{0 \, (m)}_\pm \rrangle = \pm \Gamma|\psi^{0 \, (-m)}_\pm \rrangle$ with Floquet components $|\psi^{0  \, (m)}_\pm \rrangle = 1/T \, \int_0^T  \, dt \, e^{i m \Omega t} |\psi^0_\pm (t)\rangle$ and integer $m$.
\\

\textit{Unperturbed system:} A generic Floquet-Sambe Hamiltonian $\mathcal{H}_d^0$ of an FTI projected onto the $|\psi^0_\pm (t) \rangle$ states is given by
\begin{align}   
\begin{split}
\mathcal{H}_d^0 =   \begin{bmatrix}
   \llangle \psi^0_+ |\mathcal{H}_d |\psi^0_+ \rrangle      & \llangle\psi^0_+ |\mathcal{H}_d |\psi^0_- \rrangle  \\
    \llangle\psi^0_- |\mathcal{H}_d |\psi^0_+ \rrangle & \llangle \psi^0_- |\mathcal{H}_d|\psi^0_- \rrangle  \\
    \end{bmatrix}
    = 
    \begin{bmatrix} 0     & \tau^0  \\
    (\tau^0)^\dagger & 0 
\end{bmatrix},
\end{split}
\label{eq:projected_H_floquet_0}
\end{align}
where \text{$\llangle \psi^0_\pm |\mathcal{H}_d |\psi^0_\pm \rrangle = \int_0^T  dt  \langle \psi^0_\pm(t) |H_d (t) - i \hbar \partial_t |\psi^0_\pm (t) \rangle = 0$} because $|\psi^0_\pm (t) \rangle = \pm \Gamma|\psi^0_\pm (-t) \rangle$ and \text{$\Gamma H_d(t) \Gamma = - H_d (-t)$}, and where we have defined $ \tau^0 = \llangle \psi^0_+ |\mathcal{H}_d  |\psi^0_- \rrangle$, implying that 
$|\tau^0|^2 = (\varepsilon^0)^2$. Now, $|\psi^0_\pm (t = 0) \rangle$ has support on a single sublattice of the array because $|\psi^0_\pm (t) \rangle = \pm \Gamma|\psi^0_\pm (-t) \rangle$. Thus, in the thermodynamic limit they correspond to two edge states localized at opposite edges of the array. Note, however, that these states are not expected to have support on only one sublattice  throughout the full time evolution. This is because the chiral symmetry relation is fulfilled only for the evolution operator at zero reference time. Formally, we write the following expression for the projected Floquet-Sambe Green's function:
\begin{align}   
\begin{split}
\mathcal{G}^0 (\omega) = (\omega^+ - \mathcal{H}^0_d - \mathcal{E}^0_L - \mathcal{E}^0_R )^{-1} =
    \begin{bmatrix} \omega^+ - \mathcal{E}^0_L     & \tau^0  \\
    (\tau^0)^\dagger & \omega^+ - \mathcal{E}^0_R
\end{bmatrix}^{-1},
\end{split}
\label{eq:projected_G_FS_0}
\end{align}
where $\mathcal{E}^0_{L/R}$ are projected Floquet-Sambe self-energies representing the coupling to the leads. Here
$\mathcal{E}^0_{L}$ ($\mathcal{E}^0_{R}$) only couples the $|\psi^0_+ (t) \rangle$ ($|\psi^0_- (t)\rangle$) state to the left (right) lead since $|\psi^0_\pm (t) \rangle$ are localized at separate edges at all times $t$ and therefore all other coupling terms are small. Within the wide-band limit approximation, $\mathcal{E}^0_L = -i \, \varGamma^0_L/2 
= \sum_j \Sigma_{L,j}  |\llangle j |\psi^0_+ \rrangle|^2$  and $\mathcal{E}^0_R = -i \, \varGamma^0_R/2 = \sum_j \Sigma_{R,j}  |\llangle j |\psi^0_- \rrangle|^2$, where, in contrast to the time-independent case, the summation over lattice sites now runs over both sublattices (cf. the remark above). It follows that the total photon-assisted transmission within the reduced space is given by
\begin{align}
\begin{split}
\mathcal{T}_{LR}^{0\, (0)}(\omega) &= \varGamma^{0}_R \varGamma^{0 \, (0)}_L |\mathcal{G}^0_{21} (\omega)|^2 \\
&= \frac{\varGamma^{0}_R \varGamma^{0 \, (0)}_L|\tau^0|^2 }{|(\omega^+ - \mathcal{E}^0_L)(\omega^+ - \mathcal{E}^0_R) - |\tau^0|^2|^{2}},
\end{split}
\end{align}
where $\varGamma^{0 \, (0)}_L$ is  $\varGamma^{(0)}_L$ projected on the left edge state $|\psi^0_+ (t) \rangle$.  At~$\omega = 0$ the expression for the total transmission reduces to 
\begin{align}   
\begin{split}
\mathcal{T}_{LR}^{0\, (0)} (0) = \frac{\varGamma^{0}_R \varGamma^{0 \, (0)}_L (\varepsilon^0)^2}{| \varGamma^0_L \varGamma^{0}_R / 4 + (\varepsilon^0)^2|^2}.
\end{split}
\label{eq:Transmission_proj_FS_0}
\end{align}
The reduced broadening function $\varGamma^{0\, (0)}_L$ is defined through $\varGamma^{0 \, (0)}_L = 2 i \mathcal{E}_L^{0 \, (0)}$ with $\mathcal{E}_L^{0 \, (0)} = \sum_j \Sigma_{L, j}  |\llangle j |\psi^{0 (0)}_+ \rrangle|^2$, where $|\psi^{0 (0)}_+ \rrangle$ is the time-independent part of the steady-state $|\psi^0_+ \rrangle$.

In the present case there is no straightforward way to analytically obtain the quasienergy $\varepsilon^0$ and/or states $|\psi^0_\pm \rrangle$ even for the simple model of the harmonically driven dimer array. To predict the size-dependence of the transmission one must resort to a numerical diagonalization of the Floquet-Sambe Hamiltonian $\mathcal{H}_d$ from which $\varepsilon^0$ and $|\psi^0_\pm \rrangle$ can then be obtained.
\\

\textit{Perturbed system:} Let us consider a general local time-periodic perturbation $p(t)$ located at the left edge of the topological insulator. In the following we describe how the the symmetries present 
in $p(t)$ influence the transmission spectrum. 

Any perturbation $p(t)$ modifies the projected Floquet-Sambe Green's function as follows:
\begin{align}   
\begin{split}
\mathcal{G}^0 (\omega) =
    \begin{bmatrix} \omega^+ - \mathcal{E}^0_L + \delta h^0_L     & (\tau^0 + \delta  \tau^0)  \\
    (\tau^0 + \delta  \tau^0)^\dagger & \omega^+ - \mathcal{E}^0_R + \delta h^0_R
\end{bmatrix}^{-1},
\end{split}
\label{eq:projected_G_FS_0_perturbed}
\end{align}
where $\delta h^0_{L/R} = 1/T \, \int_0^T  dt  \langle \psi^0_{\pm}(t) |p (t) |\psi^0_{\pm} (t) \rangle$ and $\delta  \tau^0 = 1/T \, \int_0^T  dt  \langle \psi^0_+(t) |p (t) |\psi^0_- (t) \rangle$. Clearly, $\delta h^0_R$ is very small because $p(t)$ is a local perturbation acting at the left edge and $|\psi^0_- (t) \rangle$ is the edge state at the right boundary.  The term $\delta  \tau^0$ may be larger but is still small because of the same reason. The only term that can potentially modify the projected transmission in a significant way is $\delta h^0_L$ with both $p(t)$ and $|\psi^0_+ (t)\rangle$ being localized at the same edge. However, perturbations satisfying the chiral relation  $\Gamma p(t) \Gamma = - p(-t)$ exactly nullify~$\delta h^0_L$. This is because $\delta h^0_L = 
1/T \, \int_0^T  dt  \langle \psi^0_{+}(t) |p (t) |\psi^0_{+} (t) \rangle = -\delta h^0_L = 0$ with the property $|\psi^0_+ (t) \rangle = \Gamma |\psi^0_+ (-t) \rangle$. In this way the transmission spectrum around zero quasienergy displays a robustness of the edge states against symmetry-preserving perturbations, including time-periodic ones.

\subsection{The pair of midgap edge states near $\hbar \Omega/2$ quasienergy}

In a finite system, the levels of two edge states that live in the anomalous gap around quasienergy $\hbar \Omega/2$ are split due to finite-size hybridization. Let us focus on one state from the pair, call it $|\psi^\pi_{I} (t) \rangle$ with quasienergy $\hbar \Omega/2 + \varepsilon^\pi$. As dictated by chiral symmetry, its complementary state $\Gamma|\psi^\pi_{I} (-t) \rangle$ then has quasienergy $-\hbar \Omega/2 - \varepsilon^\pi$. Shifting the quasienergy of $\Gamma|\psi^\pi_{I} (-t) \rangle$ by $\hbar\Omega$ we obtain the second steady state of the pair, 
$|\psi^\pi_{II} (t) \rangle = e^{i \Omega t} \Gamma|\psi^\pi_{I} (-t) \rangle$ with quasienergy $\hbar \Omega/2 - \varepsilon^\pi$. We can now project the transport problem onto the space spanned by these states. It is here convenient to use a rotated basis, $|\psi^\pi_\pm (t) \rangle = (|\psi^\pi_{I} (t)\rangle \pm |\psi^\pi_{II} (t)\rangle)/\sqrt{2}$, with the property $|\psi^\pi_\pm (t) \rangle = \pm e^{i\Omega t} \Gamma |\psi^\pi_\pm (-t) \rangle$. In terms of Floquet components, $|\psi^{\pi \, (m) }_\pm \rrangle = 1/T \int_0^T  \, dt \, e^{i m \Omega t} |\psi^\pi_\pm (t) \rangle$, this relation reads as $|\psi^{\pi \, (m)}_\pm \rangle = \pm\Gamma |\psi^{\pi \, (-m-1)}_\pm \rrangle$ with integer~$m$.
\\

\textit{Unperturbed system:} The Floquet-Sambe Hamiltonian of a generic FTI projected onto the states $|\psi^\pi_\pm (t) \rangle$ reads as 
\begin{align}   
\begin{split}
\mathcal{H}_d^\pi =  \begin{bmatrix}
   \llangle \psi^\pi_+ |\mathcal{H}_d|\psi^\pi_+ \rrangle      & \llangle\psi^\pi_+ |\mathcal{H}_d|\psi^\pi_- \rrangle  \\
    \llangle\psi^\pi_- |\mathcal{H}_d|\psi^\pi_+ \rrangle & \llangle\psi^\pi_- |\mathcal{H}_d|\psi^\pi_- \rrangle \\
    \end{bmatrix}
    = 
    \begin{bmatrix} \hbar \Omega/2     & \tau^\pi \\
    (\tau^\pi)^\dagger & \hbar \Omega/2 
\end{bmatrix},
\end{split}
\label{eq:projected_H_floquet_pi}
\end{align}
where $\llangle \psi^\pi_\pm |\mathcal{H}_d |\psi^\pi_\pm \rrangle = 1/T \, \int_0^T  dt  \langle \psi^\pi_\pm(t) |H_d (t) - i \hbar \partial_t |\psi^\pi_\pm (t) \rangle$ $ = \hbar \Omega/2$ because $|\psi^\pi_\pm (t) \rangle = \pm e^{i \Omega t}\Gamma|\psi^\pi_\pm (-t) \rangle$ and $\Gamma H_d(t) \Gamma = - H_d (-t)$. By definition, $(\varepsilon^\pi)^2 = |\tau^\pi|^2$, where $\tau^\pi= 1/T \, \int_0^T  dt \langle\psi^\pi_+(t) |H_d(t)|\psi^\pi_-(t) \rangle$. The chiral symmetry relation $|\psi^\pi_\pm (t) \rangle = \pm e^{i\Omega t} \Gamma |\psi^\pi_\pm (-t) \rangle$ requires the states to be non-zero only on a single sublattice at $t=0$, and therefore to be localized at separate boundaries in the thermodynamic limit. 
We can then express the Floquet-Sambe Green's function projected onto the two edge states as 
\begin{align}   
\begin{split}
\mathcal{G}^\pi (\omega) &= (\omega^+ - \mathcal{H}^\pi_d - \mathcal{E}^\pi_L - \mathcal{E}^\pi_R )^{-1} \\
&=
    \begin{bmatrix} \omega^+ - \mathcal{E}^\pi_L - \hbar \Omega/2     & \tau^\pi  \\
    (\tau^\pi)^\dagger & \omega^+ - \mathcal{E}^\pi_R - \hbar \Omega/2 
\end{bmatrix}^{-1},
\end{split}
\label{eq:projected_G_FS_pi}
\end{align}
where, similar to the case in Appendix D.1, $\mathcal{E}^\pi_{L/R}$ are projected Floquet-Sambe self-energies representing the coupling to the leads. The projected self-energy $\mathcal{E}^\pi_{L}$ ($\mathcal{E}^\pi_{R}$) only connects $|\psi^\pi_+ (t) \rangle$ ($|\psi^\pi_- (t)\rangle$) to the left (right) lead because $|\psi^\pi_\pm (t) \rangle$ are each separately localized at one of the edges at any time $t$. Using the same notation as before, the total photon-assisted transmission within the reduced space can then be written as 
\begin{align}
\begin{split}
\mathcal{T}_{LR}^{\pi \, (0)}(\omega) &= \varGamma^\pi_R \varGamma^{\pi \, (0)}_L |\mathcal{G}^\pi_{21} (\omega)|^2 \\
&= \frac{\varGamma^\pi_R \varGamma^{\pi \, (0)}_L |\tau^\pi|^2}{|(\omega^+\! -\! \mathcal{E}^\pi_L \!+ \!\hbar \Omega/2)(\omega^+ \!-\! \mathcal{E}^\pi_R \!+\! \hbar\Omega/2) \!- \!|\tau^\pi|^2|^{2}}.  
\end{split}
\end{align}

We are particularly interested in the transmission at $\text{$\omega = \hbar\Omega/2$}$:
\begin{align}   
\begin{split}
\mathcal{T}_{LR}^{\pi \, (0)} (\hbar\Omega/2) = \frac{\varGamma^\pi_R \varGamma^{\pi \, (0)}_L  (\varepsilon^\pi)^2}{| \varGamma^\pi_L \varGamma^{\pi}_R / 4 +  (\varepsilon^\pi)^2|^2},
\end{split}
\label{eq:Transmission_proj_FS_pi}
\end{align}
where  $\mathcal{E}_L^\pi = -i\varGamma^\pi_L/2$ and $\mathcal{E}_L^\pi =  -i \varGamma^\pi_R/2$ from the wide-band limit approximation. Note that Eq. (\ref{eq:Transmission_proj_FS_pi}) is the anomalous-gap analogue of Eq. (\ref{eq:Transmission_proj_FS_0}) derived in Appendix D.1. Here the projected Floquet-Sambe self-energies are given by $\mathcal{E}^\pi_L = \sum_j \Sigma_{L,j}  |\llangle j |\psi^\pi_+ \rrangle|^2$  and $\mathcal{E}^\pi_R = \sum_j \Sigma_{R,j}  |\llangle j |\psi^\pi_- \rrangle|^2$, where, like in the previous subsection, the sum over lattice sites runs over both sublattices. The reduced projected broadening function reads as $\varGamma^{\pi \, (0)}_L = 2 i \mathcal{E}_L^{\pi \, (0)}$ with $\mathcal{E}_L^{\pi \, (0)} = \sum_j \Sigma_{L, j}  |\llangle j|\psi^{\pi (0)}_+ \rrangle|^2$, where $|\psi^{\pi (0)}_+ \rrangle$ is the time-independent part of the steady-state $|\psi^\pi_+ \rrangle$. Recall that the Floquet components of the edge state at $\varepsilon = \hbar \Omega / 2$ satisfy $|\psi^{\pi \, (m)}_\pm \rangle = \pm\Gamma |\psi^{\pi \, (-m - 1)}_\pm \rrangle$ leading to $\varGamma_L^{\pi \, (0)} \leq \varGamma_L^{\pi}/2$. In accordance to  Eq.~(\ref{eq:Transmission_proj_FS_pi}) this means that $\mathcal{T}_{LR}^{\pi \, (0)} (\hbar\Omega/2) \leq 1/2$.
 
 As in Appendix~D.1, to predict the dependence of the transmission spectrum on the size of the dimer array we perform a separate numerical computation and retrieve from it the parameters $\varepsilon^\pi$ and $|\psi^\pi_\pm\rrangle$.
\\

\textit{Perturbed system:} Let us take a general local time-periodic perturbation $p(t)$ localized at the left edge of the topological insulator. The projected Floquet-Sambe's function is then modified as follows:
\begin{align}   
\begin{split}
\mathcal{G}^\pi (\omega) =
    \begin{bmatrix} \omega^+ - \mathcal{E}^\pi_L -\hbar \Omega/2 + \delta h^\pi_L     & (\tau^\pi + \delta  \tau^\pi)  \\
    (\tau^\pi + \delta  \tau^\pi)^\dagger & \omega^+ - \mathcal{E}^\pi_R - \hbar \Omega/2 +\delta h^\pi_R
\end{bmatrix}^{-1},
\end{split}
\label{eq:projected_G_FS_pi_perturbed}
\end{align}
with $\delta h^\pi_{L/R} = 1/T \, \int_0^T  dt  \langle \psi^\pi_{\pm}(t) |p (t) |\psi^\pi_{\pm} (t) \rangle$ and $\delta  \tau^\pi = 1/T \, \int_0^T  dt  \langle \psi^\pi_+(t) |p (t) |\psi^\pi_- (t) \rangle$. The terms $\delta h^\pi_R$ and $\delta  \tau^\pi$ are small because they contain a product of $p(t)$ and $|\psi^\pi_- (t) \rangle$, each localized at a separate edge of the system. The remaining term  $\delta h^\pi_L$ may potentially influence the projected transmission. However, if the perturbation fulfills the chiral symmetry condition $\Gamma p(t) \Gamma = - p(-t)$ we get $\delta h^\pi_L = 1/T \, \int_0^T  dt  \langle \psi^\pi_{+}(t) |p (t) |\psi^\pi_{+} (t) \rangle = -\delta h^\pi_L = 0$ by applying the relation $|\psi^\pi_+ (t) \rangle = e^{i \Omega t} \Gamma |\psi^\pi_+ (-t) \rangle$. It follows that the effect on the transmission peak at $\hbar \Omega/2$ from a local perturbation which respects chiral symmetry is strongly suppressed.

Our numeric results in Fig.~9 suggest that time-periodic symmetry-breaking perturbations seem to have significantly larger effect on the topological transmission peak at $\omega = \hbar \Omega/2$ than at $\omega = 0$. The localization of the edge states at both band gaps is approximately the same since the transmission peaks are maximum at approximately the same system sizes, so this can not be the reason for such severe difference in responses to the symmetry breakage. The reason must be in the intrinsic difference between these states. 
Recall that a symmetry-protected edge state at $\varepsilon = \hbar \Omega/2$ has the same probability weights correspoding to $m$ and \text{$(-m-1)$} Floquet indices, in contrast to a symmetry-protected edge state at $\varepsilon = 0$ with equal probability weights correspoding to $m$ and $(-m)$ Floquet indices. Exactly this property, along with the fact that Floquet states usually decay very rapidly with the Floquet index $m$~\cite{Rudner}, implies that a time-periodic symmetry-breaking perturbation shall have a significantly larger effect on the transmission peak at $\omega = \hbar \Omega /2$ than at $\omega = 0$: The product $\int_0^T \, dt \, \langle \psi^\pi_+(t) |e^{i n \Omega t} |\psi^\pi_+ (t) \rangle$ is generically larger than $\int_0^T \, dt \, \langle \psi^0_+(t) |e^{i n \Omega t} |\psi^0_+ (t) \rangle$ for integer $n\geq 1$.


\end{document}